\documentclass[twocolumn,showpacs,preprintnumbers,amsmath,amssymb]{revtex4}

\usepackage{graphicx}
\usepackage{dcolumn}
\usepackage{bm}
\usepackage{epstopdf}

\begin{document}
\title{Local elasticity map and plasticity in a model Lennard-Jones glass.}

\author{Michel Tsamados$^1$}
\email{michel.tsamados@lpmcn.univ-lyon1.fr}
\author{Anne Tanguy$^1$}
\author{Chay Goldenberg$^1$}
\author{Jean-Louis Barrat$^1$}

\affiliation{$^1$Universit\'e de Lyon; Univ. Lyon I,  Laboratoire de
Physique de la Mati\`ere Condens\'ee et des Nanostructures; CNRS,
UMR 5586, 43 Bvd. du 11 Nov. 1918, 69622 Villeurbanne Cedex,
France}

\date{\today}
\setcounter{page}{1}

\begin{abstract}
In this work we calculate the local elastic moduli in a weakly polydisperse 2D Lennard-Jones glass
undergoing a  quasistatic  shear deformation at zero temperature.
The numerical method uses coarse grained microscopic expressions for the strain,
displacement and stress fields. This method allows us to calculate the local elasticity tensor and to quantify  the deviation from linear elasticity (local Hooke's law) at different coarse-graining  scales. From the results a clear picture emerges of an amorphous material with strongly spatially heterogeneous elastic moduli that simultaneously satisfies Hooke's law at scales larger than a characteristic length scale of the order of five interatomic distances. At this  scale the glass appears as a composite material composed of a rigid scaffolding and of soft zones. Only recently calculated in non homogeneous materials, the local elastic structure plays a crucial role in the elasto-plastic response of the amorphous material.
  For a small macroscopic shear strain the structures associated with the non-affine displacement field appear directly related to the spatial structure of the elastic moduli. Moreover for a larger macroscopic shear strain we show that zones of low shear modulus concentrate most of the strain in form of plastic rearrangements.
  The spatio-temporal evolution of this local elasticity map and its connection with long term dynamical heterogeneity as well as with the plasticity in the material is quantified. The possibility to use this local  parameter as a predictor of subsequent local plastic activity is also discussed.

\end{abstract}

\maketitle

\newpage

\section{Introduction}

It is commonly acknowledged that the mechanical properties and the rheology of a wide class of amorphous glassy materials involve localized structural rearrangements of the order of five interatomic distances involving of the order of 100 particles in three dimensions or 20 in two dimensions~\cite{Tanguy2006,Mayr2006}. These events are frequently compared with the rearrangements that take place in aging glassy materials where local events are thermally activated~\cite{Nandagopal2003,Malandro1998}.
It has been proposed that these rearrangements can organise during a mechanical deformation, through a cascade mechanism to form shear bands~\cite{Maloney2006,Maloney2004,Langer2001,Bailey2007,Demkowicz2005a}.
This phenomenon of strain localization has been observed experimentally in alloys, metallic glasses, polymers, granular media, foams and colloids~\cite{Debregeas2001,Lauridsen2002,Varnik2004} as well as in numerous simulations, both of model systems such as Lennard-Jones glasses~\cite{Varnik2003,Shi2006,Shi2005,Tanguy2006}, as in more realistic simulations ~\cite{Delogu2008,Delogu2008a,Delogu2008b,Bailey2004}. On a theoretical level various current models of the rheology of glasses predict reasonably well the macroscopic mechanical properties of these materials ~\cite{Falk1998,Falk1999,Argon1995,Bulatov1994,Bulatov1994a,Bulatov1994b,Sollich1997,
   Sollich1998,Baret2002,Picard2005,Rottler2003,Wyart2005b,Demkowicz2005,Maloney2004,Maloney2006,Lemaitre2007a,Malandro}.
These models generally involve the consideration of zones with prescribed elasto-plastic properties,
whose microscopic identification remains however elusive. So despite a recent and rich literature concerning
the connection between the structure of the glass, the intrinsic dynamics of these irreversible events and the dynamical heterogeneity in amorphous
 systems~\cite{Argon2006,Demkowicz2005,Demkowicz2005a,Demkowicz2004,Shi2006,Shi2005,Shi2005a,
 Capaldi2002,Downton2008,Kustanovich2000,Toninelli2005,Widmer-Cooper2006,Widmer-Cooper2004,Doliwa2003a,Doliwa2003}
questions such as what type of, where and how local plastic rearrangement occur in a deformed glass remain unanswered.

A common general idea is that the deformation will take place in `weak' zones, somehow characterized by abnormally low elastic constants and increased mobility. This picture of the glass as composed of a patchwork of `rigid' relatively strongly bonded (but amorphous) domains separated by `soft' regions (walls) has long been postulated \cite{Alexander1998} and is at the heart of many theoretical models of the glass transition\cite{Stillinger1988}. Moreover within this framework it appears tempting to relate the dynamical heterogeneities observed in glassy materials near and below the glass transition temperature to the spatially inhomogeneous elastic constant network. Experimental evidence of the heterogeneous glass structure is shown in \cite{Vidal} and models \cite{Schirmacher1998,Taraskin2001} including fluctuating elastic moduli have been introduced in the last two decades to describe some acoustical (boson peak \cite{Duval2002}) and thermal (specific heat anomalies \cite{Heat1,Heat2,Heat3}) properties of glasses.
In order to check such assumptions, and attempt to relate structure and dynamics in amorphous materials a new route was recently opened by calculating local elastic constants in such heterogeneous material~\cite{Yoshimoto2004}.
 Simulation methods for the calculation of the elastic constants of solids can be classified in two groups. One can either use equilibrium fluctuation formulas, as described in \cite{Yoshimoto2004} or perform explicit deformations and derive these constants from the stress-strain response of the material. In this paper we implement the deformation  approach in a framework of continuum mechanics developed in \cite{Goldhirsch2002}, that extends the applicability of classical continuum mechanics to smaller scales. While this method requires several computations to calculate the different elastic constants, it remains simple and accurate, and has the advantage to give a systematic  estimate  of the deviation from linear elasticity as a function of the coarse graining scale.

  Section \ref{methods} presents the system under study, a well characterized
  two dimensional Lennard-Jones system, and the method used to obtain the local elastic moduli.
  The results for the moduli and the resulting picture of the glass structure are described in section \ref{moduli}.
  Finally, in section \ref{plastic}, we show how the local elastic inhomogeneities are connected to the plastic activity as our system undergoes quasistatic shear deformation, and the relation between local elastic moduli and local dynamics is also explored.

\section{Methods}
\label{methods}

\subsection{Sample preparation}

The systems studied throughout this paper are slightly polydisperse two-dimensional Lennard-Jones glasses, described in detail in ~\cite{Tanguy2002}.
 The interaction energy between two particles is 
 $4 \epsilon ((\frac{\sigma_i+\sigma_j}{2r})^12-(\frac{\sigma_i+\sigma_j}{2r})^6)$ with $\sigma_i$ the diameter of particle i.
 Our typical sample contains 10 000 particles in a square simulation box, with a polydispersity of $5\%$ on the size of the particles.
All  numerical values in the following are expressed in Lennard-Jones units (LJU), where the average particle diameter and the interaction energy are equal to unity. The corresponding density of the glass $\rho = 0.925 LJU$ was chosen to minimize the initial pressure of the quenched amorphous solid to $P = 0.25$.
The quenching procedure, from the liquid state to well below the glass transition temperature,
 that produces the initial configurations consists in a sequence of thermal steps -
 at temperatures T = 1.0, 0.5, 0.1, 0.05, 0.01, 0.005 and 0.001 - each lasting 100 LJU and
 thermostated through simple velocity rescaling. This quench is followed by an energy minimization
 scheme to bring the system at a local energy minimum and at zero temperature, this protocol is also described in detail in ~\cite{Tanguy2002}.
Two protocols were used to shear the material.
  In protocol one, two layers of particles with thickness $2\, LJU$ are singled out and assumed to constitute parallel solid walls that will impose the deformation to the system. The resulting cell
  has a thickness  $Ly=100$ and a width $Lx=104$. The configuration is submitted to
  shear by applying constant displacement steps $\delta u_{x}$ to the particles of the upper wall, corresponding to an elementary strain of about $\epsilon=5.10^{-5}$. Between two displacement steps the entire system is relaxed with fixed walls into its nearest equilibrium position.
  The total strain of the sample under these rigid walls boundary conditions is $\epsilon_{total} = 1.65$ (i.e. $165 \%$ deformation).
  In protocol two the system of thickness $Ly=104$ and width $Lx=104$ is sheared with Lees-Edwards
  periodic boundary conditions. The elementary strain step during the shear is here $\epsilon=2.5 10^{-5}$. The total strain under these Lees-Edwards boundary conditions is $\epsilon_{total} = 0.5$ (i.e. $50 \%$ deformation). To check possible size effects a series of glasses of different sizes (containing up to 250 000 particles) were produced under the same quenching procedure and analyzed in the very early linear domain ($\epsilon_{total} < 10^{-6}$). 
  
\begin{figure}[!hbtp]
\begin{center}
\includegraphics*[width=8cm]{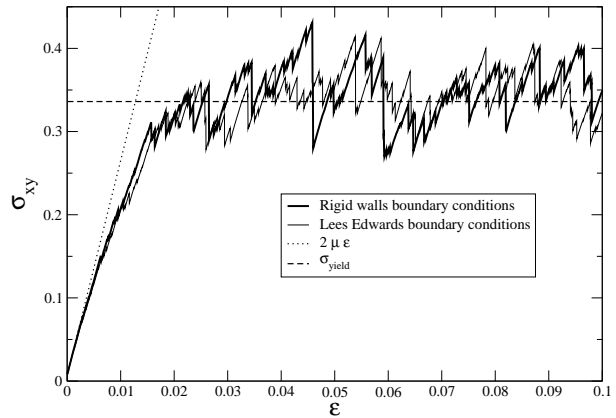}
\caption{Stress-strain mechanical response of the 2D polydisperse Lennard-Jones model glass under shear, indicating the separation between an elastic loading phase and a plastic flow behavior. Rigid (protocol one) and Lees Edwards (protocol two) boundary conditions are respectively represented in thick and thin black lines.}
\label{fig1}
\end{center}
\end{figure}

  In the following section we derive the coarse-
  grained (CG) method used to calculate the local elastic tensor $C$ as well as other local
  CG fields such as the strain tensor $\epsilon$, the stress tensor $\sigma$, the density $\rho$.
  This method is applied to  configurations in the transient regime $\epsilon < 2.5 10^{-2}$ as well
  as in the fully developed plastic regime $\epsilon > 2.5 10^{-2}$. The domains analyzed are shown in figure \ref{fig1}.

\subsection{Measuring local elastic constants}

As stated in the introduction,  various methods have been proposed in the literature for the numerical calculation of elastic constants. The elastic constants are defined \cite{Barron1965} as the coefficients $C_{\alpha\beta\gamma\delta}$ of the second order expansion of the strain-energy density as a function of the local Green-Lagrange strain components
\begin{equation}
\label{eqn:W}
\frac{W-W_o}{V}=C^o_{\alpha\beta}\epsilon_{\alpha\beta}+\frac{1}{2}C_{\alpha\beta\gamma\delta}\epsilon_{\alpha\beta}\epsilon_{\gamma\delta}+...,
\end{equation}
or as the first order expansion of the local stress components $\sigma_{\alpha\beta}$ as a function of the local linear strain \cite{Salencon2001}
\begin{equation}
\label{hooke}
\sigma_{\alpha\beta}=C^o_{\alpha\beta}+C'_{\alpha\beta\gamma\delta}\epsilon^{lin}_{\gamma\delta}+...
\end{equation}
with
$$\sigma_{\alpha\beta}=\frac{1}{V}\frac{\partial W}{\partial \epsilon^{lin}_{\alpha\beta}}$$ This second definition can be seen as a linear local fit of the energy and is known as Hooke's law.
In case of an unstressed solid, the two definitions are strictly equivalent, but in case of a solid with initial stresses ($C^o\ne 0$ as in our Lennard-Jones glasses), the difference between the two components $C_{\alpha\beta\gamma\delta}$ and $C'_{\alpha\beta\gamma\delta}$ depends on the quenched stress components, due to the non-linear dependence of $\epsilon$ as a function of $\epsilon^{lin}$ \cite{Barron1965}. In our case, we have checked that the contribution of quenched stresses can indeed be neglected \cite{Tsamados2009}, but this formal difference must be mentioned.

At finite temperature one can use a fluctuation type method to determine (\ref{eqn:W}), and for a system at equilibrium the elastic modulus tensor and the stress fluctuations are related through the following fluctuation formula \cite{Ray1985,Lutsko1988,Lutsko1989}:
\begin{eqnarray}
\label{eqn:c:xi}
&C_{\alpha\beta\gamma\delta} =
C_{\alpha\beta\gamma\delta}^{Born}+2Nk_{B}T(\delta_{\alpha\beta}\delta_{\beta\delta}+\delta_{\alpha\delta}\delta_{\beta\gamma})-\nonumber\\
&\frac{V_{0}}{k_{B}T}\left[\langle\hat t_{\alpha\beta} \hat t_{\gamma\delta}\rangle-\langle\hat t_{\alpha\beta}\rangle\langle \hat t_{\gamma\delta}\rangle \right].
\end{eqnarray}

The term in square brackets (here the brackets denote a thermal average) is called the fluctuation term of the microscopic stress tensor $\hat t_{\gamma\delta}$. The Born term $C_{\alpha\beta\gamma\delta}^{Born}$ corresponds to the instantaneous elastic modulus for a sample under an uniform and infinitesimal strain. Equation \ref{eqn:c:xi} has been used for a long time to compute elastic constants of amorphous materials \cite{Roux1988} but only recently applied to the calculation of \textit{local  }elastic constants in amorphous polymers ~\cite{Yoshimoto2004}, in metallic glasses \cite{Mayr2009}, in composites ~\cite{Papakonstantopoulos2008} as well as in small samples of Lennard-Jones glasses~\cite{Ilyin2008} above the glass transition temperature.
If one is interested, on the other hand, in the mechanical properties of a material at zero temperature one must use the zero temperature limit of equation \ref{eqn:c:xi}, that was shown in \cite{Lutsko1988} to be
\begin{equation}
\label{eqn:c:xi2}
C_{\alpha\beta\gamma\delta} =
C_{\alpha\beta\gamma\delta}^{Born}-{\vec\Xi}_{\alpha\beta}.{\bf H}^{-1}.{\vec\Xi}_{\gamma\delta}.
\end{equation}
Here
\begin{equation}
\label{eqn:xi:cauchy}
{\vec\Xi}_{i,\gamma\delta}
\equiv\left.\frac{\partial\sigma_{\gamma\delta}}{\partial\vec r_{i}}\right|_{\eta\to0}
\quad.
\end{equation}
can be understood as the forces which would result from an elementary homogeneous deformation of all the particles in the strain direction $\eta_{\gamma\delta}$ and where {\bf H} is the dynamical matrix of second derivatives of the potential energy with respect to atomic positions.
At zero temperature, the `relaxation-fluctuation' term (second term on the right hand side of equation \ref{eqn:c:xi2}) was shown \cite{Tanguy2002} to account for an important fraction of the absolute value of elastic constants in amorphous systems (this is also the case in crystals with a complex unit cell). This failure of the Born term alone (first term on the right hand side of equation \ref{eqn:c:xi}) to accurately describe the mechanical properties of the material can be traced back to the existence of a non-affine deformation field, which stores part of the elastic energy. Unfortunately, the direct evaluation of the `relaxation-fluctuation' term is not straightforward as it necessitates the inversion of the Hessian matrix. Hence the actual procedure to accurately determine the elastic constants of athermal systems generally consists in carrying out explicitly an affine deformation of all coordinates and a corresponding deformation of the simulation cell, then letting the atomic positions relax towards the nearest energy minimum within the deformed cell. One calculates the corresponding local stress increments and linear strain components, and can subsequently recover the elastic tensor coefficients by solving Hooke's law \ref{hooke}.

In this paper we propose to extend this method to the calculation of the local elastic constants at different scales of coarse graining. Following~\cite{Goldhirsch2002}, we present first the coarse-grained expressions that we use to measure locally the stress and strain fields, we then pursue with the derivation of the local elastic tensor. \\

\textbf{Local strain tensor}\\

The simplest and crudest approach is the so called Voigt assumption where one assumes that the strain is homogeneous in the sample.
While this approach is exact for simple unit cell crystals it cannot be valid for disordered materials and leads to wrong estimates of the elastic constants as discussed above. It was proposed in refs \cite{Falk1998} as an improvement on the mean field assumption, that the local strain should be determined from  a best fit of  the actual particle displacements within a small subsystem to those generated by an adjustable strain tensor. Other possibilities for defining a local strain include the consideration of a dyadic tensor built on the links between neighboring particles \cite{Graner2008}. In this study we will use an alternative definition proposed in \cite{Goldhirsch2002}, according to which  the (linear) strain tensor is written in terms of a coarse grained displacement field:
\begin{equation}
\epsilon^{\mathrm{{lin}}}_{\alpha\beta}({\bf r},t)=\frac{1}{2}\bigg(\frac{\partial
 u_\alpha ^{\mathrm{{lin}}}({\bf r},t)}{\partial r_\beta}
+\frac{\partial
 u_\beta ^{\mathrm{{lin}}}({\bf r},t)}{\partial r_\alpha}\bigg)
\end{equation}
where the superscript `lin' denotes the linear order in the displacements. The linear displacement ${\bf u}^{\mathrm{{lin}}}({\bf r},t)$ is defined as the linear dependence on the displacement of the individual particles, of the temporal integration of the coarse-grained velocity. The latest is computed by the way of the momentum density whose coarse-grained expression satisfies the mass conservation equation. As shown in ref \cite{Goldhirsch2002}, by integration of the coarse-grained velocity field $\dot {\bf u}({\bf r},t)$, one obtains  the following expression for the linear order in displacement:
\begin{equation}
{\bf u}^{\mathrm{{lin}}}({\bf r},t) =
\frac{\sum_i m_i {\bf u}_i(t) \phi\left[{\bf r}-{\bf r}_i(t)\right]}
{\sum_j m_j  \phi\left[{\bf r}-{\bf r}_j(t)\right]} \label{linearu}
\end{equation}\\
where $\phi({\bf r})$ is the coarse-grained function, for example as in all calculations done in this paper, a gaussian of width $W$. ${\bf u}^{\mathrm{{lin}}}({\bf r},t)$ is a continuous and differentiable function of space that allows to compute $\epsilon^{\mathrm{{lin}}}$ by spatial derivation. The diff\'erence ${\bf u}^{\mathrm{{lin}}}({\bf r}_i,t)-{\bf u}_i(t)$, where ${\bf u}_i$ is the actual displacement of the particle $i$, is the fluctuating part of the displacement \cite{Goldenberg2007}. It can not be negligible when the displacement of the particles is strongly varying with space, as it is shown at very small scales in our Lennard-Jones glasses \cite{Goldenberg2007}.\\

\textbf{Local stress tensor}\\

The most commonly used definition is the atomic stress introduced by Irving Kirkwood \cite{Irving1950} as
\begin{equation}
\sigma_{\alpha\beta, i}\equiv -\frac{1}{V_{i}}\sum_{j} f_{i/j\alpha}(t) r_{ij\beta}(t)
\end{equation}
$V_i$ being the volume of the Voronoi cell associated with atom i, and $f_{i/j\alpha}$ the force exerted by atom j on atom i in the direction $\alpha$.
If the velocity term is neglected this formula when averaged over a sufficiently large sample corresponds to the the Cauchy stress (which  measures the actual mechanical force per unit area), at smaller length scales though it does not, strictly speaking,  verify the  momentum conservation equation.

An expression for the stress components that satisfies the momentum conservation in the framework of the previous coarse-graining has been obtained in \cite{Goldhirsch2002}

\begin{eqnarray}
\label{sigma2}
\sigma_{\alpha\beta}({\bf r},t) =&\\
 -\frac{1}{2} \sum_{i}\sum_{j(\neq i)} &f_{i/j\alpha}(t) r_{ij\beta}(t) \int_0^1 ds \phi\left[ {\bf r}-{\bf r}_i(t)+s {\bf r}_{ij}(t)\right].\nonumber
\end{eqnarray}
with $\phi$ the coarse-graining function. This expression has also been used recently in \cite{Cormier2001}.\\

\textbf{Derivation of linear elasticity - local elastic tensor}\\

Now having calculated the stress and strain tensors locally we are in a position to derive the corresponding components of the local elastic tensor as in the case of a macroscopic deformation.
This presuposes that equation \ref{hooke} remains valid at a local level and one can therefore apply the same symetry arguments as in the macroscopic case to reduce the number of independent coefficients of the elasticity tensor. Since the linear strain is symetric by definition, and the stress is symmetric in the absence of torques there are at most 9 constants in 2D. As discussed in \cite{Barron1965,Salencon2001} the existence of a strain energy function from which the equations of elasticity can be derived by variational methods implies a further symetry of the elastic tensor :
$C_{\alpha\beta\gamma\delta} = C_{\gamma\delta\alpha\beta}$, reducing the number of independent thermodynamic stiffness to 6 in 2D.
To extract the 6 independent elastic coefficients necessitates at least two independent deformation modes on our sample. Each deformation provides three linear equations for the moduli. The general stress-strain relation in terms of matrices is written as follow, using a Voigt type notation. For each coarse-graining scales $W$, the coarse-grained stress and strain components are measured on a grid, and are expressed respectively as the $3*1$ column vectors $\hat T$ and $\hat E$ and one has in 2D the relation $\hat T=\hat{\bf C}\hat E$ that means:
\begin{equation}
\label{eqn:C2D}
\left(
\begin{array}{c}
\sigma_{xx} \\
\sigma_{yy} \\
\sqrt{2}\sigma_{xy} \\
\end{array}
\right)
=
\left(
\begin{array}{ccc}
C_{xxxx} & C_{xxyy} & C_{xxxy}\\
C_{xxyy} & C_{yyyy} & C_{yyxy}\\
C_{xxxy} & C_{yyxy} & C_{xyxy}
\end{array}
\right)
\left(
\begin{array}{c}
\epsilon_{xx} \\
\epsilon_{yy} \\
\sqrt{2}\epsilon_{xy} \\
\end{array}
\right)
\end{equation}
This expression can be compared with the expression obtained in the framework of homogeneous and isotropic linear elasticity:

\begin{equation}
\label{eqn:Ciso}
\left(
\begin{array}{c}
\sigma_{xx} \\
\sigma_{yy} \\
\sqrt{2}\sigma_{xy} \\
\end{array}
\right)
=
\left(
\begin{array}{ccc}
\lambda+2\mu & \lambda & 0\\
\lambda & \lambda+2\mu & 0\\
0 & 0 & 2\mu
\end{array}
\right)
\left(
\begin{array}{c}
\epsilon_{xx} \\
\epsilon_{yy} \\
\sqrt{2}\epsilon_{xy} \\
\end{array}
\right)
\end{equation}
where $\mu$ is the shear modulus and $\lambda$ the Lam\'e coefficient (($\lambda+\mu$) is the inverse of the compressibility modulus in $2D$).

The use of 2 such deformations therefore closes the system of unknowns, giving 6 equations for 6 unknowns. Nevertheless, in order to estimate the deviation from linear elasticity the numerical procedure used here consists in applying three different uniform deformation modes, two uniaxial stretching parallel to the x and y axis and a simple shear parallel to the x axis. This procedure provides 9 linear equations for the moduli.
The stress components which are not used in this procedure are then calculated using these elastic moduli, and their values compared to those computed directly using equation \ref{sigma2}. As a measure of the extent to which the system is described by linear elasticity at a given position and for a given value of the coarse-graining scale $W$ we use the root mean square, $\Delta$, of the relative differences between the stress components calculated by employing the measured moduli and the directly measured exact values (normalized by the norm of the exact values).

For each configuration we calculate also the 3 eigenvalues $c_{i}$ and eigenvectors $E_{i}$ for $i=1,2,3 $ of the local tensor $\hat C$.
The comparison between equations \ref{eqn:C2D} and \ref{eqn:Ciso} would provide $c_1=2\mu$, $c_2=2\mu$ and $c_3=2(\lambda+\mu)$ in an homogeneous and isotropic system.
We will now discuss the results obtained in our model Lennard-Jones glass, as a function of the coarse-graining scale $W$.

\section{Analysis of the local moduli}
\label{moduli}

\begin{figure}[!hbtp]
\begin{center}
\includegraphics*[width = 8cm]{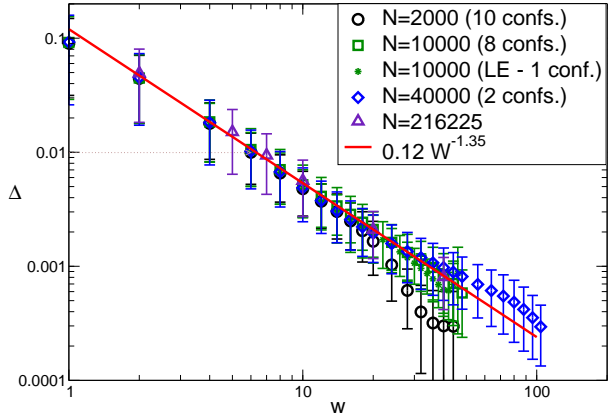}
\caption{Deviation $\Delta$ from linear elasticity as a function of the coarse-graining parameter W. For $W \ge 5$ Hooke's law is satisfied locally with more than $1\%$ accuracy.}
\label{fig2}
\end{center}
\end{figure}

\begin{figure}[!hbtp]
\begin{center}
\includegraphics*[width = 8cm]{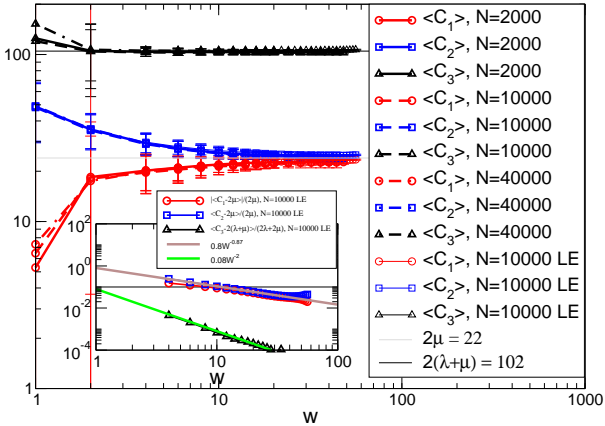}
\caption{The three eigenvalues of the elasticity tensor averaged spatially as a function of the coarse-grain length W. Inset: log-log plot of the convergence to the limit value obtained by a coarse graining on the whole system size.}
\label{fig3}
\end{center}
\end{figure}

\begin{figure}[!hbtp]
\begin{center}
\includegraphics*[width = 8cm]{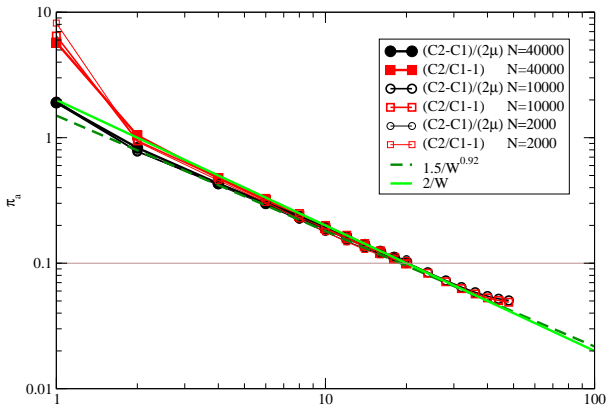}
\caption{ Anisotropy parameter $\pi_a$ as a function of the coarse-graining scale $W$.}
\label{fig4}
\end{center}
\end{figure}

The domain of validity of  Hookes's law can now be measured by the $W$-dependence of the error function $\Delta$. It is shown in figure \ref{fig2}.
This figure shows that the error $\Delta$ goes progressively to zero, thus validating  Hooke's law at large coarse-graining scales $W$. However, this convergence obeys a power-law $\Delta\propto W^{-1.32}$ and therefore doesn't exhibit a characteristic scale above which Hooke's law would be true. Nevertheless, we can see in figure \ref{fig2} that the error $\Delta$ is already less than $1\%$ for $W\ge 5$. It means that above $W=5$, the error made in computing the stress components using the elastic modulus is less that a hundred of the actual value of the stress, that is far smaller than the contribution of quenched stresses for example. We can thus consider that the system obeys Hooke's law  reasonably well on scales larger than $W=5$.
Below that scale, different factors could explain the departure from Hooke's law: the first one is the coarse-graining. Below $W=5$ it has been shown in a previous paper \cite{Goldenberg2007} that the contribution of the coarse-graining deformation to the actual one is small. The contribution of the fluctuating field can not be neglected, giving rise to high values of the real strain and therefore a significant decrease of the elastic moduli. In fact, the fluctuating field is not differentiable, so that it is  not possible in this case to compute quantitatively the linear strain components. It is one of the interests of the use of a coarse-graining field to deal with differentiable displacements fields. The fluctuating field appears thus as a `noise field'. An additional strongly fluctuating term should then be taken into account for $W<5$ in an attempt to describe accurately the mechanical behavior of the material. Another contribution to the departure from Hookes's law at small scale is due to the coupling to second and third neighbors outside the volume element. This contribution (not taken into account here since we are restricted to first order derivatives in the displacement field) could be introduced by considering the contribution of higher order derivatives of the deformation in the framework of linear but long-range elasticity \cite{NonLocalBook}. This paper is devoted to the measurement of elastic moduli, thus we will leave these considerations for further studies, and focus now on the computation of elastic moduli in the domain of validity, that is for $W>5$.
\\

Figure \ref{fig3} shows the average value of each of the eigenvalues $<\overline{c_{i}}>$ of $\hat C$ as a function of the coarse-graining scale $W$. The notation $\overline c$ stands for a spatial averaging over the sample and angular brackets stand for a statistical averaging over different configurations. The spatial averaging is obtained by computing the elastic moduli on a grid with elements of width W/2. The number of the different samples used in the statistical averaging - for each given size $L$ - is the same as indicated in the figure \ref{fig2}. We get first the average value of the $(L/(W/2))^2$ values obtained on the grid, for each sample. Then we average the values obtained on the different samples.
The figure \ref{fig3} shows a progressive convergence for large W to the values obtained in the framework of homogeneous and isotropic linear elasticity. Indeed, for very large $W$, $<\overline{c_1}>$ and $<\overline{c_2}>$ go to the same value $2\mu\approx 22$ obtained also by looking at the macroscopic response of the system to various mechanical sollicitations, and $<\overline{c_3}>$ converges to twice the inverse compressibility $2(\lambda+\mu)\approx 102$ measured as well by the global response of the sample \cite{Tanguy2002,Leonforte2004}. The method used here to compute the elastic moduli of the system for large $W$ appears thus to be consistent  with measurements of the global response of the system. 

The convergence of the average local elastic moduli to their macroscopic value is independent of the system lateral size $L$  as long as  $W<0.5 L$. For larger $W$, the boundary conditions (Lees Edwards or fixed walls) may affect the convergence and cause finite size effects. As shown in the inset of figure \ref{fig3}, the moduli decay approximately as $1/W^\alpha$ to their limit value, with $\alpha\approx 0.87$ for $c_1$ and $c_2$, and $\alpha\approx 2$ for $c_3$. The inverse compressibility converges thus far more quickly to its macroscopic value, than the shear moduli. The difference between the macroscopic value and the spatial average of the small scale measurements of $c_i$ ($2\mu\approx 22$ while $<\overline{c_1}(W=5)>\approx 18$ for example) is due to the inhomogeneous strain field. By looking at $(<C>-C_{\infty})/C_{\infty}$ (inset of figure \ref{fig3}), it appears that for $W>5$, the discrepancy to the macroscopic value is already less than $1\%$ for $<c_3>$, while it becomes less than $10\%$ only for $W>10$, for $<c_1>$ and $<c_2>$. We conclude that for $5<W<10$ the moduli are well defined, but the measured values are not compatible with homogeneous elasticity since the different moduli involved at different coarse-graining scales have different scale dependence. We did not find any solid explanation for the non-trivial power-law appearing in this convergence. It appears that the convergence of the inverse compressibility is inversely proportional to the volume $W^2$, and the convergence of the shear moduli closer to a surface effect $\propto W$. Note also that one of the shear moduli is smaller than the limiting value, while the other is larger. This difference between smooth and hard directions will now allow us to quantify  the anisotropy of the local mechanical response.
\\

The anisotropy measured at small scale can be quantified by the ratio $(<\overline{c_2}>-<\overline{c_1}>)/(2\mu)$ that goes to zero for large $W$. We call it the anisotropy parameter $\pi_a$. It is shown in figure \ref{fig4}. It can be noticed that $c_1$ and $c_2$ always obey $<\overline{c_1}> \leq 2\mu$ and $<\overline{c_2}> \geq 2\mu$, so $\pi_a \geq 0$.
The decay of the anisotropy parameter $\pi_a$ (figure \ref{fig4}) obeys also a power-law $\propto 1/W^{0.92}$, close to $1/W$. These power-law decays prevent us to define properly a characteristic scale above which the homogeneous and isotropic behaviour is recovered.
In figure~\ref{fig4}, we see that the anisotropy parameter becomes less than $10\%$ for $W>20$ only. It means that, below $W=20$, it is possible to find locally a well defined direction associated with a very low local shear modulus.
At this scale, the anisotropy in the mechanical response cannot be neglected.

\begin{figure}[!hbtp]
\begin{center}
\includegraphics*[width = 8cm]{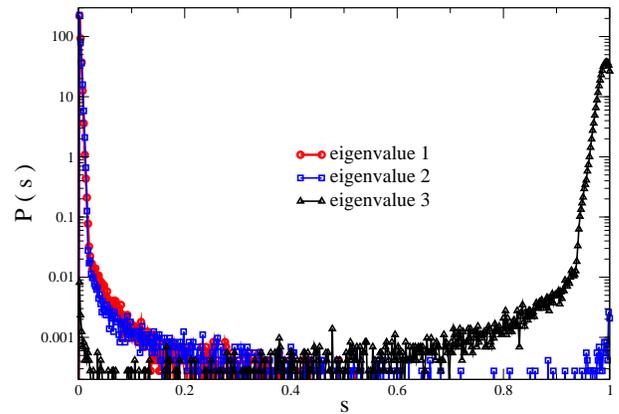}
\caption{Distribution for each eigenvalue $E_{i}$ of the local elasticity tensor of the quantity $S_{i}=\left(tr(E_{i})\right)^{2}/2tr(E
_{i}^2)$. The distribution is obtained over the entire plastic flow regime.
}
\label{fig5}
\end{center}
\end{figure}

The preferred direction of strain is given locally by the analysis of the eigenvectors $E_{1}$, $E_{2}$ and $E_{3}$. Each eigenvector contains the 3 distinct elements of a 2D strain tensor whose eigendirections $\textbf{e}_{1}$ and $\textbf{e}_{2}$ ($\textbf{e}_{1}$ and $\textbf{e}_{2}$ are orthogonal) are computed.
We plot on figure \ref{fig5} the distribution of the local quantity $S_{i}=\left(tr(E_{i})\right)^{2}/2tr(E_{i}^2)$, that takes the value 0 if the deformation is pure shear and 1 for pure dilatation. One observes as expected that the two deformations associated with the two lowest eigenvalues are of pure shear type while the third deformation is a pure compression.

\begin{figure}[!hbtp]
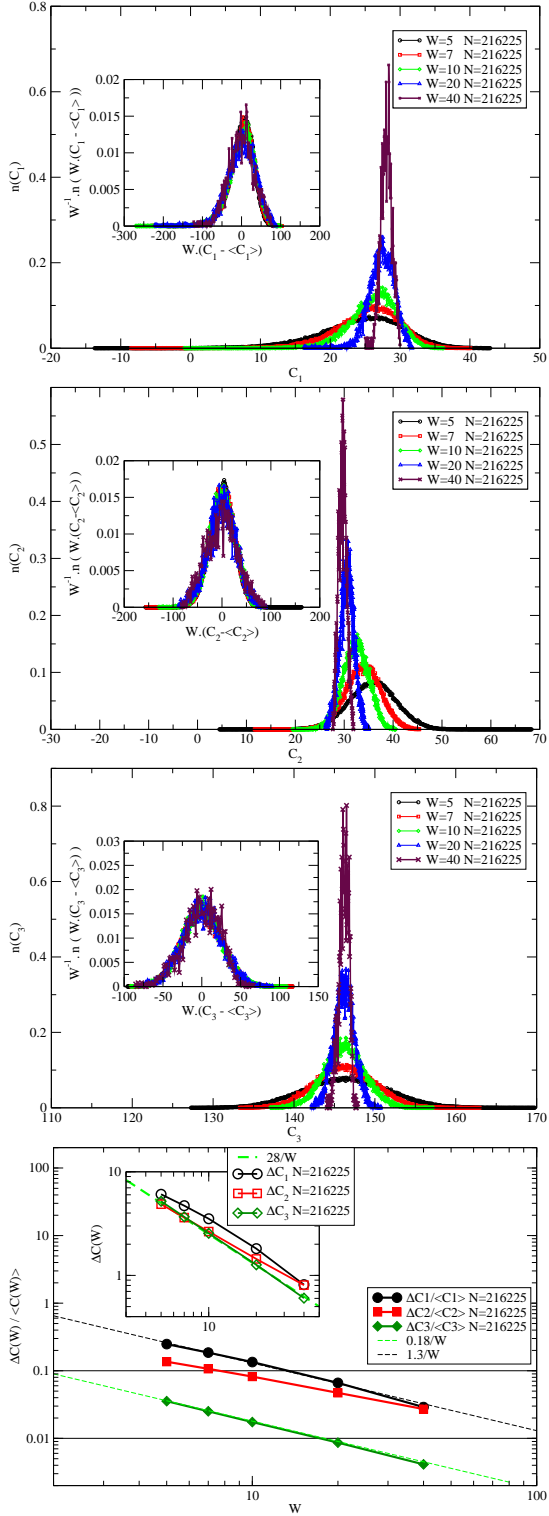

\begin{center}
\includegraphics*[width = 0.4\textwidth]{figure6a.eps}
\includegraphics*[width = 0.4\textwidth]{figure6b.eps}
\includegraphics*[width = 0.4\textwidth]{figure6c.eps}
\includegraphics*[width = 0.4\textwidth]{figure6d.eps}
\caption{The figures (a), (b) and (c) show, respectively, the distributions of the three lowest eigenvalues $c_{1}$, $c_{2}$ and $c_{3}$ for different coarse-graining lengths $W$. The inset of each figure show the distributions rescaled by $W$. Figure (d) shows that the distributions get narrower as $W$ increases to converge to the macroscopic values of the elastic constants. The variances of the distributions is plotted as a function of $W$. The variances normalized by the average values of the distributions are also plotted. Both curves show a power law evolution with $W$.}
\label{fig6}
\end{center}
\end{figure}

In order to explore more deeply the inhomogeneities of the elastic moduli inside the system, we will now study their distribution as a function of $W$, and then their spatial correlations.
The distributions of $c_1$, $c_2$ and $c_3$ are shown in figure \ref{fig6} for various $W$ and for $N=216225$.
First, we can see in these distributions, that zones with negative moduli can appear if the coarse-graining scale $W$ is sufficiently small, as already observed in \cite{Yoshimoto2004}. It is not in contradiction with the mechanical stability of our system, since the local elastic moduli computed here are only part of the second order derivative of the total mechanical energy, due to the coarse-graining, as well as to the non trivial dependence of the non-affine local deformation as a function of the applied displacement.
The rescaling of the distributions by $W$ is also shown on the figures \ref{fig6}. It is very good for sufficiently large values of $W$ (typically $W>10$).
The variance of the distribution as a function of $W$ is shown in the inset of figure \ref{fig6}d. It decays as $1/W$ and is always smaller than the corresponding average value. It is thus impossible to identify a characteristic lengthscale by the comparison of the variance and the average value of the moduli $c_i$.

The decay of the relative fluctuations $\Delta c/<\overline c>\propto 1/W$ for a given $W$ can be interpreted in the framework of a sum of uncorrelated variables with finite variance, the distribution being nearly gaussian. The apparent rescaling of the distribution thus corresponds to a sum of spatially uncorrelated variables.
Note that, while this ratio is very small for $c_3$ for every value of $W$ ($\Delta c_3/c_3 < 0.01$ for $W\ge 5$), it is much larger for $c_1$ ($\Delta c_1/c_1 > 10\%$ while $W\leq 15$ for $N=216225$). We can thus conclude that the inhomogeneity is far more pronounced in the shear modulus $c_1$ than in the compressibility $c_3$. The inhomogeneity of $c_1$-modulus is even far from been negligible while $W\leq 15$.

By comparing this result with the result obtained for the isotropy, we can conclude that at a scale $W>20$ the system becomes reasonably isotropic and homogeneous. Below this scale, it is homogeneous but not yet isotropic for $15<W<20$. All these results are summarized in Table \ref{Table}. Of course, these values must be nuanced by the fact that the criteria used are a little arbitrary and simply related with a comparison of orders of magnitude. No characteristic lengthscale can be clearly identified since all the quantities checked here have a power-law dependence with the scaling length $W$.

\newcommand\T{\rule{0pt}{4.5ex}}
\newcommand\B{\rule[-2.5ex]{0pt}{0pt}}
\begin{center}
\begin{table}[h!b!p!]
\begin{tabular}{ | p{2.5cm} | p{1cm} | p{1cm} | p{1cm} | p{1cm} | p{1cm} |}
\hline 
W \T \B & 0 & 5 & 10 & 15 & 20 \\ \hline 
Hooke's law \T \B & NO & YES & YES & YES & YES \\ \hline
Homogeneity \T \B  &  &  &  &  &  \\
\begin{LARGE}$\frac{\langle\overline{c}\rangle-2\mu}{2\mu}$\end{LARGE}$<10\%$ \T \B & NO & NO & YES & YES & YES \\
\begin{LARGE}$\frac{\Delta c}{\langle\overline{c}\rangle}$\end{LARGE}$<10\%$ \T \B & NO & NO & NO & YES & YES \\ \hline
Isotropy \T \B &  &  &  &  &    \\
\begin{LARGE}$\frac{c_{2}-c_{1}}{2\mu}$\end{LARGE}$<10\%$ \T \B & NO & NO & NO & NO & YES\\ \hline
\end{tabular}
\caption{Analysis at different coarse-graining length scales W.}
\label{Table}
\end{table}
\end{center}

\begin{figure}[!hbtp]
\begin{center}
\includegraphics*[width = 8cm]{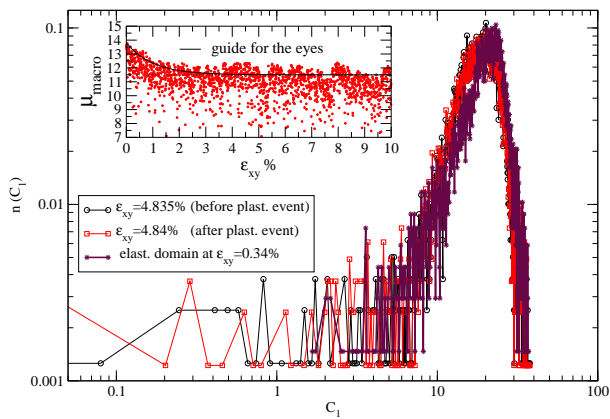}
\caption{Distributions of the local shear modulus $c_{1}$ averaged over the elastic regime and for two configurations in the plastic regime, just before and just after a rearrangement. \textbf{Inset}: Evolution with the macroscopic strain of the instantaneous macroscopic shear modulus defined as $\Delta\sigma_{xy,macro}=\mu_{macro}\Delta\epsilon_{xy,macro}$.}
\label{fig8}
\end{center}
\end{figure}

\begin{figure}[!hbtp]
\begin{center}
\includegraphics*[width = 8cm]{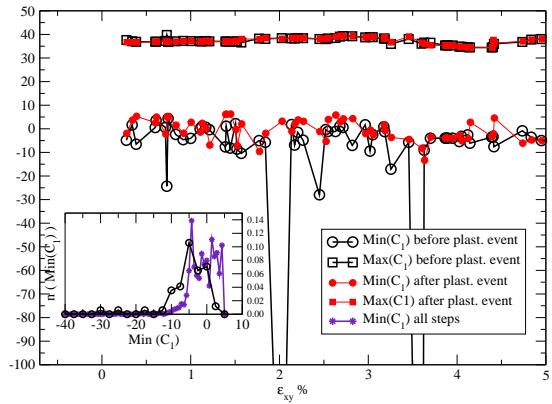}
\caption{Evolution with the strain of the maximal and minimal values of $c_{1}$.
\textbf{Inset}: Distribution of the maximal and minimal values of the local elastic modulus $c_{1}$ over the entire plastic regime for all configurations, and for configurations just before the occurence of a plastic event.}
\label{fig9}
\end{center}
\end{figure}

The distribution of the elastic moduli has been checked along the full deformation process. We can thus compare the distribution of moduli during elastic deformation and during plastic flow. We can also compare these distributions before and after a plastic rearrangement occurred in the system. The plastic rearrangements are identified here as in \cite{Tanguy2006,Tsamados2008} by the decrease of the total stress as a function of the applied strain. We see in figure \ref{fig8} that the distribution is progressively displaced to smaller values of the shear modulus, before it reaches a plastic plateau, but remains globally unchanged during all the plastic flow.
During the plastic flow, the difference between the distributions appears on extremal values. Before a plastic rearrangement occurs, the smallest value of $c_1$ (open circle in figure \ref{fig9}) is generally smaller than after the event occurred (full circle). The distribution of the smallest value of this elastic modulus is shown in the inset of figure \ref{fig9} for the full deformation process. It confirms that the smallest values are smaller before a plastic rearrangement than after. We will come back to this observation in section \ref{plastic}.

\begin{figure}[!hbtp]
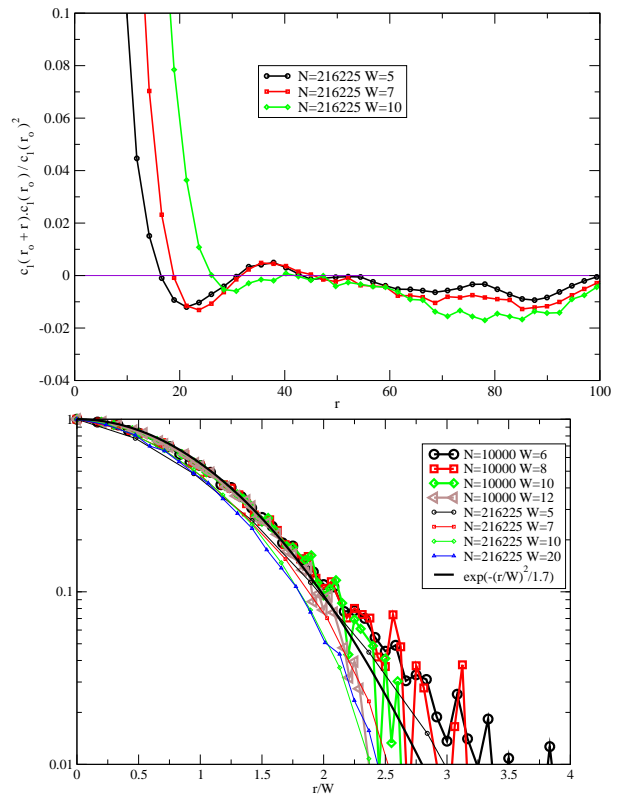

\begin{center}
\includegraphics*[width = 8cm]{figure10a.eps}
\includegraphics*[width = 7cm]{figure10b.eps}
\caption{(a) Spatial correlation function of the local shear modulus $c_{1}$: $\langle (c_{1}(0)-\overline{c_{1}})(c_{1}(r)-\overline{c_{1}}) \rangle$. (b) Scaling of the function with the coarse-graining length $W$ for small values of r. }
\label{fig10}
\end{center}
\end{figure}

Finally, it can be interesting to analyse in details the spatial correlations of the moduli in our systems. The spatial correlations of the lowest modulus $c_1$ are shown in figure \ref{fig10} for various $W$ and $N=216225$. The spatial correlations go to zero at large distances. It shows spatial oscillations with very small amplitude, that are visible while $W<10$, but that disappear for $W\ge 10$. The distance between successive maxima is about a few tens interatomic distances, but seems to be size-dependent. Unfortunately, our data are not sufficiently precise to allow us to characterize this size-dependence. For $r<3W$ only, the spatial correlations are controlled by the $W$-dependence of the coarse-graining function (see figure \ref{fig10}(b)). It can be fitted by a gaussian $\propto exp(-(r/W)^2/1.7)$. We can thus conclude that the spatial correlations are dominated by the $W$-dependence of the coarse-graining function at small distances, but displays oscillations at larger scales. The domain for which these oscillations are visible ($W<10$) corresponds to the domain in which the heterogeneity in the distribution of the moduli is noticeable. All these results show that a coarse-graining at scales $W>>10$ will loose informations (on the heterogeneities, on the spatial correlations, even on the local anisotropy.). In the future we will thus use only the value $W=5$ for the coarse-graining scale.

\section{Structure and dynamics}
\label{plastic}

\subsection{Structural relaxation}

In the previous section we have shown how the system could be decomposed into regions of different elastic stiffness. We now discuss how this elastic heterogeneity is related with the `dynamics' of the system undergoing quasistatic, plastic shear deformation.
To this end, we obtained the  local elastic parameter $c_1$ for configurations of the sheared system separated by an incremental strain of $\Delta\epsilon\simeq 5.10^{-5}$, during a set of intervals each within a total strain of $\sim10\%$ in protocol one, under rigid boundary conditions and $\sim6\%$ in protocol two, under Lees Edwards boundary conditions. The coarse-graining scale $W=5$ was chosen in all the subsequent analysis as the limit of applicability of linear elasticity. We recall that at this scale, $<\overline{c_1}>(W=5)\approx 18$.

\begin{figure}
\includegraphics[width=8cm]{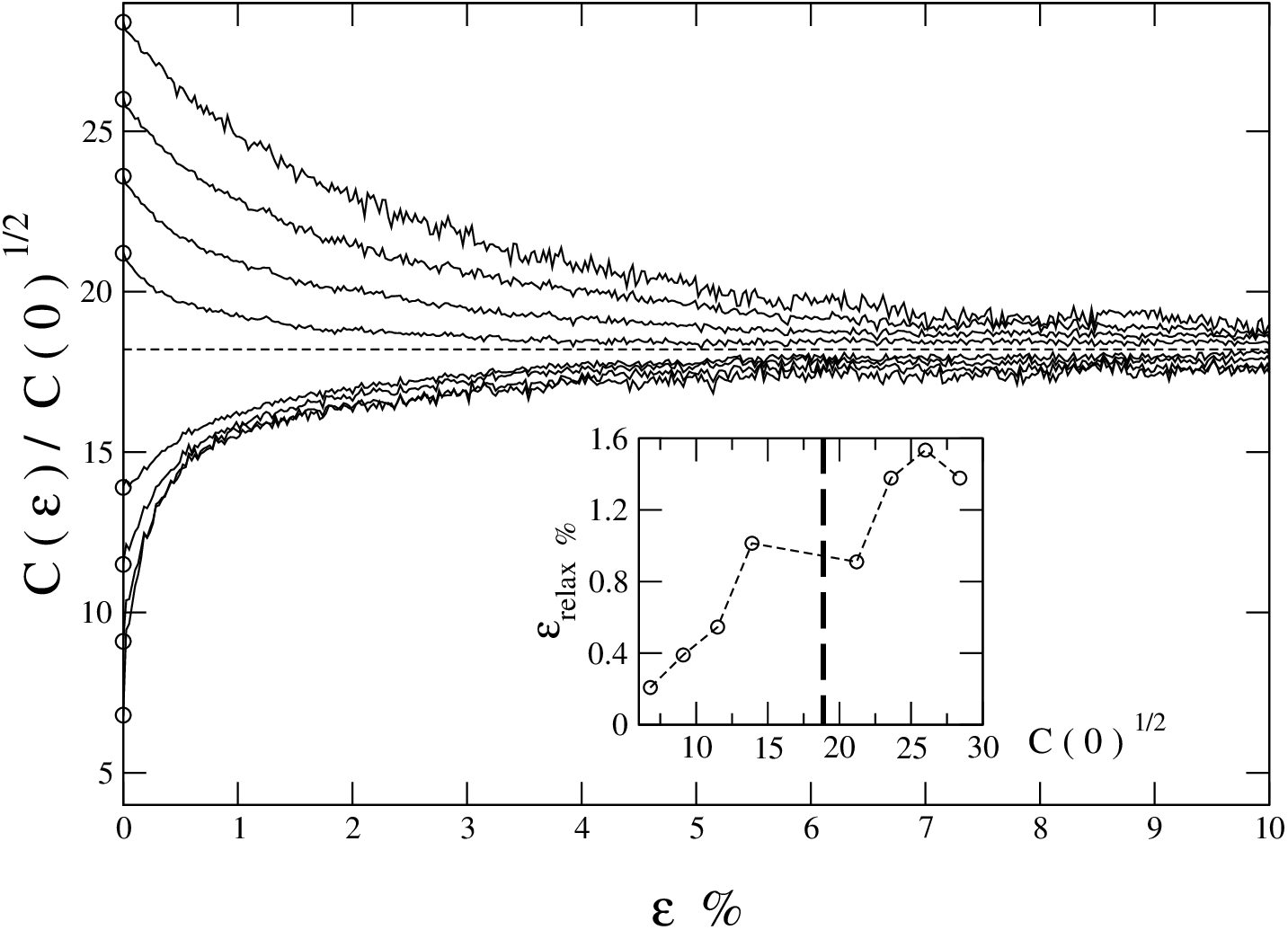}
\caption{Autocorrelation function $C(\Delta\epsilon)=\langle \overline{c_{1}({\bf r},\epsilon+\Delta\epsilon)c_{1}({\bf r},\epsilon)}\rangle$ of the order parameter conditionned by the value of $c_{1}({\bf r},\epsilon)$ at the origin. The curves correspond from bottom to top to increasing values of $c_{1}({\bf r},\epsilon)$ at the origin taken in the ranges [5-7,5],[7.5,10],[10,12.5],12.5,15],[20-22.5],[22.5,25],[25,27.5],and [27.5,30] as marked by the circles. The dashed line corresponds to $\overline{c_1}$. Inset: the typical relaxation strain $\epsilon_{relax}$ associated with  each rigidity $c_{1}\sim C(0)^{1/2}$ defined as the strain for which the rescaled autocorrelation function has decayed by half.}
\label{fig11}
\end{figure}

One can first quantify the global relaxation time (strain) associated with the field $c_{1}$ by calculating the spatially averaged two point correlation function ${\mathcal C}(\Delta\epsilon)=\langle \overline{c_{1}({\bf r},\epsilon+\Delta\epsilon)c_{1}({\bf r},\epsilon)}\rangle$ where the notation $\overline A$ stands for a spatial average over the sample and angular brackets stand for a statistical average over the strain origins $\epsilon$. In order to relate the relaxation strain to the local rigidity of the material we also calculate the two-point autocorrelation function of the shear modulus conditioned by its  initial value. For each sampled rigidity $c_{1}$ we plot the rescaled autocorrelation function ${\mathcal C}(\Delta\epsilon)/{\mathcal C}(0)^{1/2}$. In figure \ref{fig11} we see that the functions ${\mathcal C}(\Delta\epsilon)/{\mathcal C}(0)^{1/2}$ tends  in the limit of large strains to the limiting value  $\overline{c_{1}}$, independently of the initial value of the shear modulus $c_{1}$. Therefore in the stationary plastic flow regime the local shear modulus $c_{1}({\bf r},\epsilon+\Delta\epsilon)$ becomes uncorrelated for sufficiently large strains $\Delta\epsilon$ from its value at the origin $c_{1}({\bf r},\epsilon)$, showing that in this model glass the local elasticity map does not phase separate into permanently rigid and soft regions but rather evolves dynamically under shear. In the inset of figure \ref{fig11} we associate with each sampled rigidity $c_{1}$ a relaxation strain $\epsilon_{relax}(c_{1})$ defined as the strain for which the rescaled autocorrelation function has decayed by half. We see that $\epsilon_{relax}(c_{1})$ is a monotonic increasing function of the local rigidity parameter, that saturates for ${\mathcal C}(0)^{1/2}>\overline{c_1}$. We see also that softer regions (${\mathcal C}(0)^{1/2}<\overline{c_1}$) relaxes more quickly than rigid ones.

\begin{figure}[!hbtp]
\begin{center}
\includegraphics*[width = 8cm]{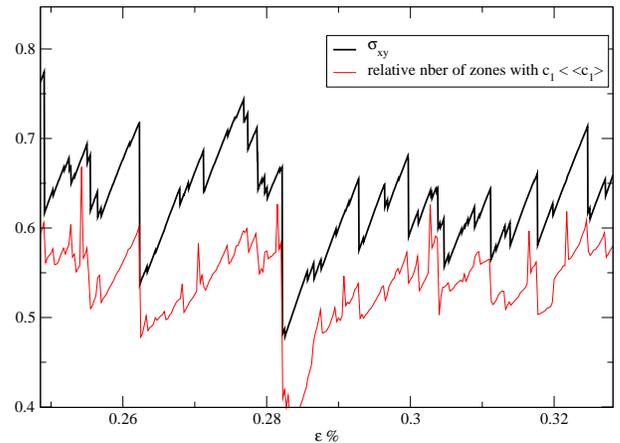}
\caption{Evolution of the total shear stress (thick line) and the relative number of soft zones (relative number of points ${\bf r}$ where $c_1({\bf r})<\overline{c_1}$) as a function of the applied shear strain, in th plastic flow regime.}
\label{fig12}
\end{center}
\end{figure}

To our knowledge this result represents one of the first numerical evidence of a clear relation between a structural order parameter and the dynamics in a glassy system. The measurement of the local elasticity map presents the advantage to be independent of the specificity of the glass under study, requiring only the measure of a local stress and strain. This result confirms the description introduced in section \ref{moduli} of the material as a composite material made of `soft' fast relaxing zones (for $c_{1}\le\overline{c_{1}}$) and of `rigid' stable zones (for $c_{1}\ge\overline{c_{1}}$). As seen in figure \ref{fig11} the strain associated with the rigid `scaffolding' of the material is found to be of the order of $\epsilon_{relax}\simeq 1.5\%$. This value is similar to the strain necessary to enter the fully plastic regime $\epsilon_{plastic}\simeq 2\%$ (see figure \ref{fig1}). As suggested in figure \ref{fig12} where the relative number of soft zones evolves in parallel with the total shear stress and reaches a maximum percentage ($\approx 60\%$) before a large plastic event, one can also see this typical strain as the necessary strain required to achieve percolation through the material of soft zones, i.e. when the material's rigid scaffolding is no longer connected \cite{Shi2005}. Figure \ref{fig13} shows the evolution of the rigid scaffolding over a total strain of $\simeq 0.9\%$.

This  relaxation strain of the order of $1.5\%$ can be compared with the typical strain separating two irreversible rearrangements \cite{Tanguy2006} in the sample $\Delta\epsilon_{event}\sim0.1\%$. An estimate of the number of plastic rearrangements required to renew the rigid `scaffolding' of the material can therefore be given as $\epsilon_{relax}/\Delta\epsilon_{event}\simeq15$, hence typically 15 events for $10000$ particles.

\subsection{Relation between the local elasticity map, the local mobility and the long time dynamical heterogeneity}

We have characterized the dynamics of the underlying structure in terms of the shear rigidity order parameter. We want now to see how this structure is coupled to the displacement field in the sheared material.

\begin{figure}
\includegraphics[width=8cm]{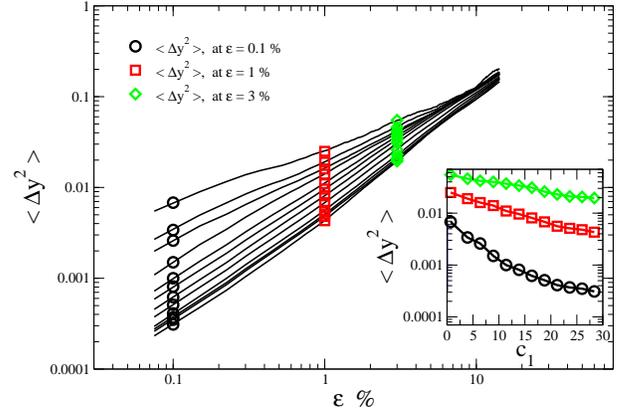}
\caption{Mean-square displacement $\langle \Delta y^{2}\rangle$ for different values of the original local shear modulus $c_{1}$. From top to bottom the curves correspond to $c_{1}$ taken in the ranges [0-2.5],[2.5,5],[5-7.5],[7.5,10],[10,12.5],12.5,15],[20-22.5],[22.5,25],[25,27.5],and [27.5,30]. Inset: $\langle\Delta y^{2}\rangle$ is shown as a function of $c_{1}$ for different relaxation strains.}
\label{fig14}
\end{figure}

\begin{figure*}
\centering
\includegraphics*[width = 0.4\textwidth]{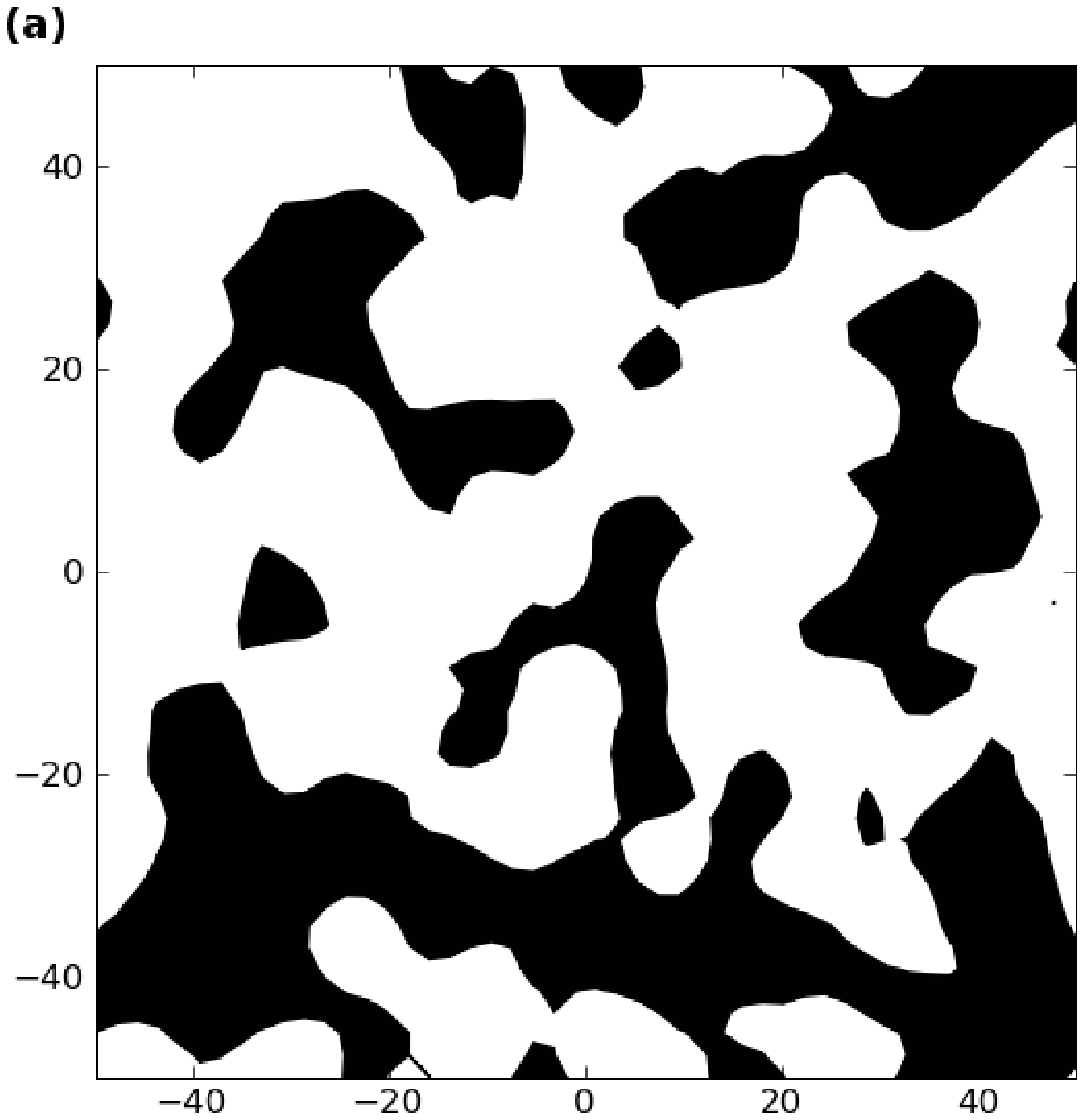}
\includegraphics*[width = 0.4\textwidth]{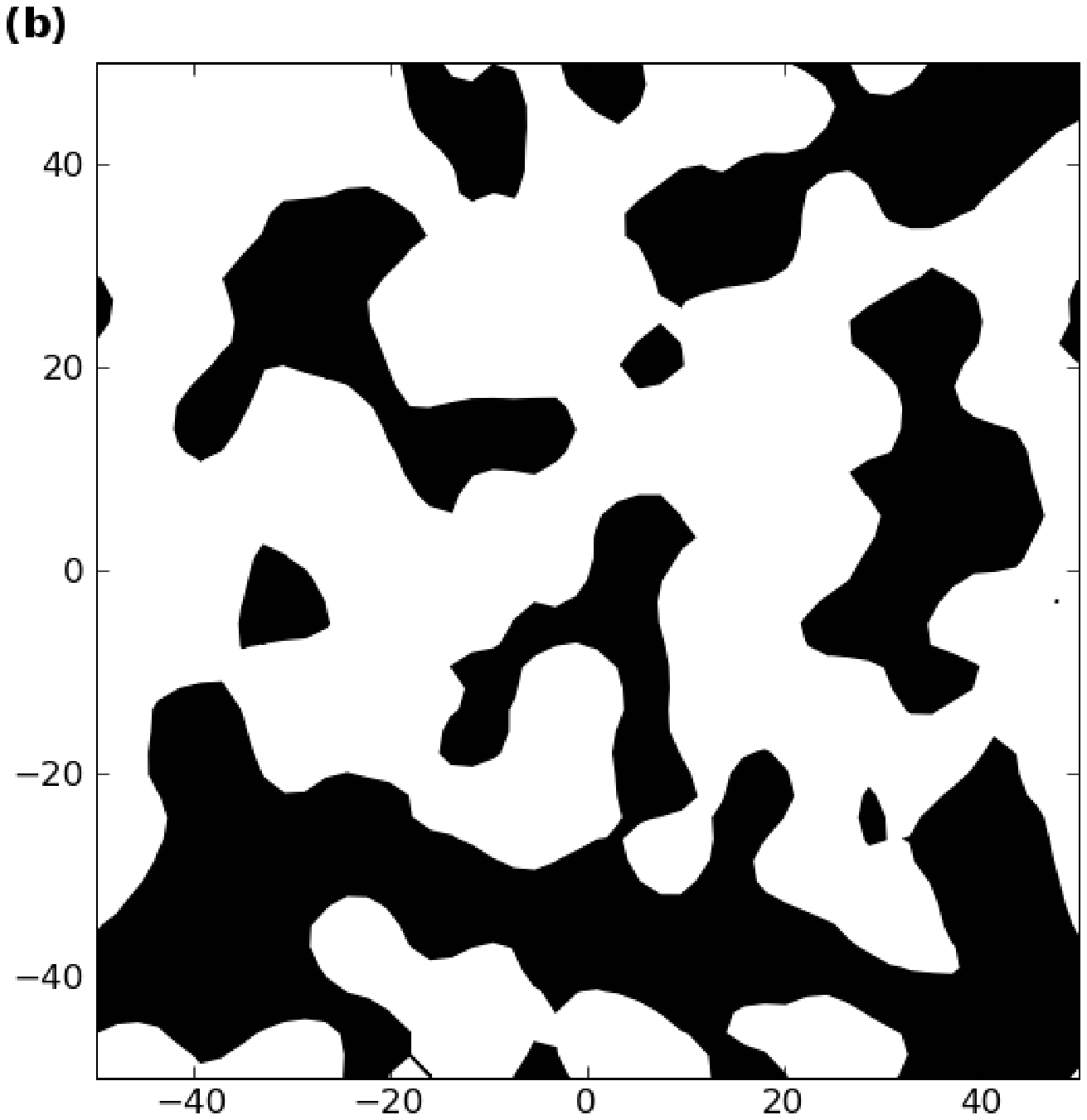}
\includegraphics*[width = 0.4\textwidth]{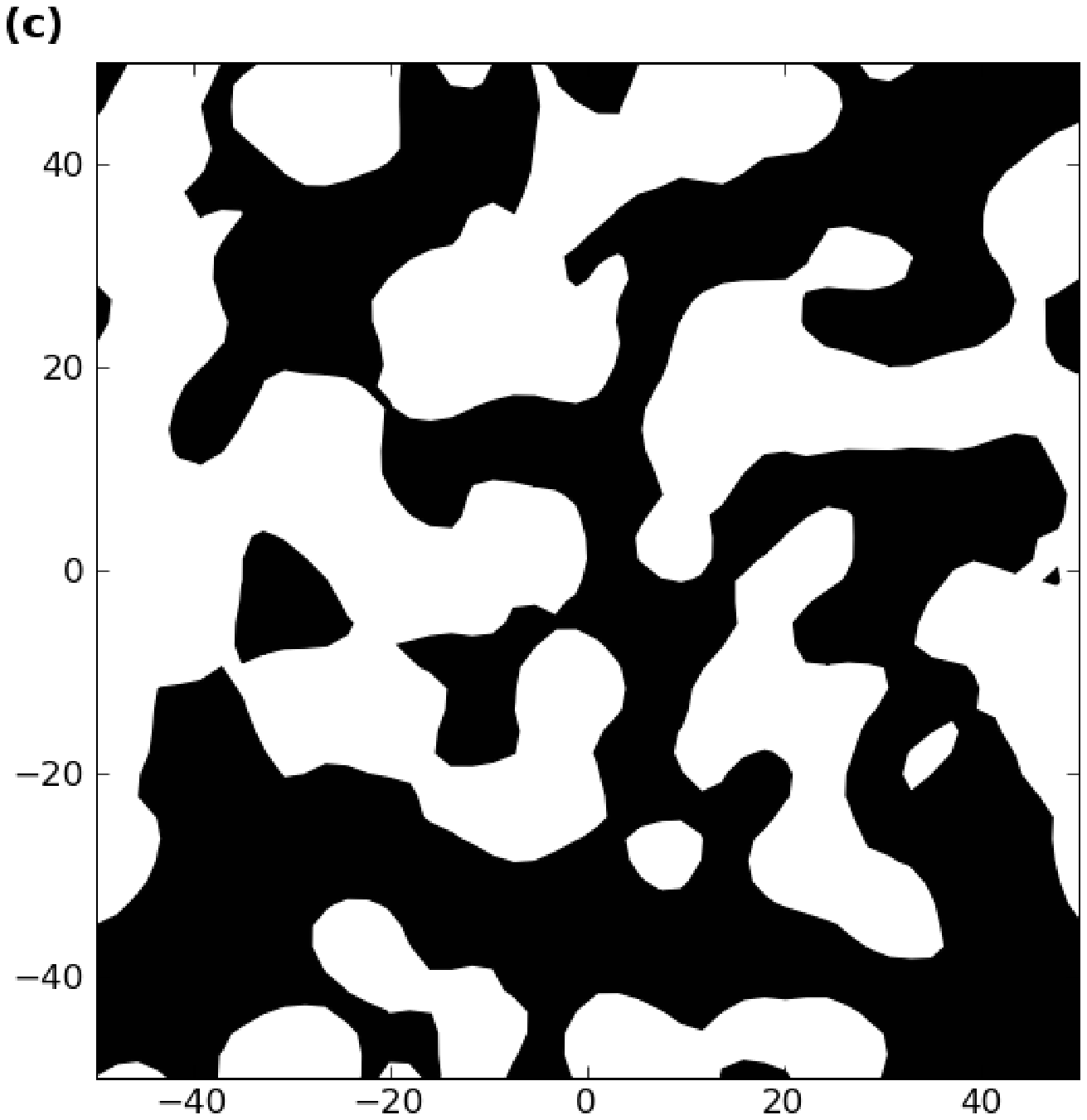}
\includegraphics*[width = 0.4\textwidth]{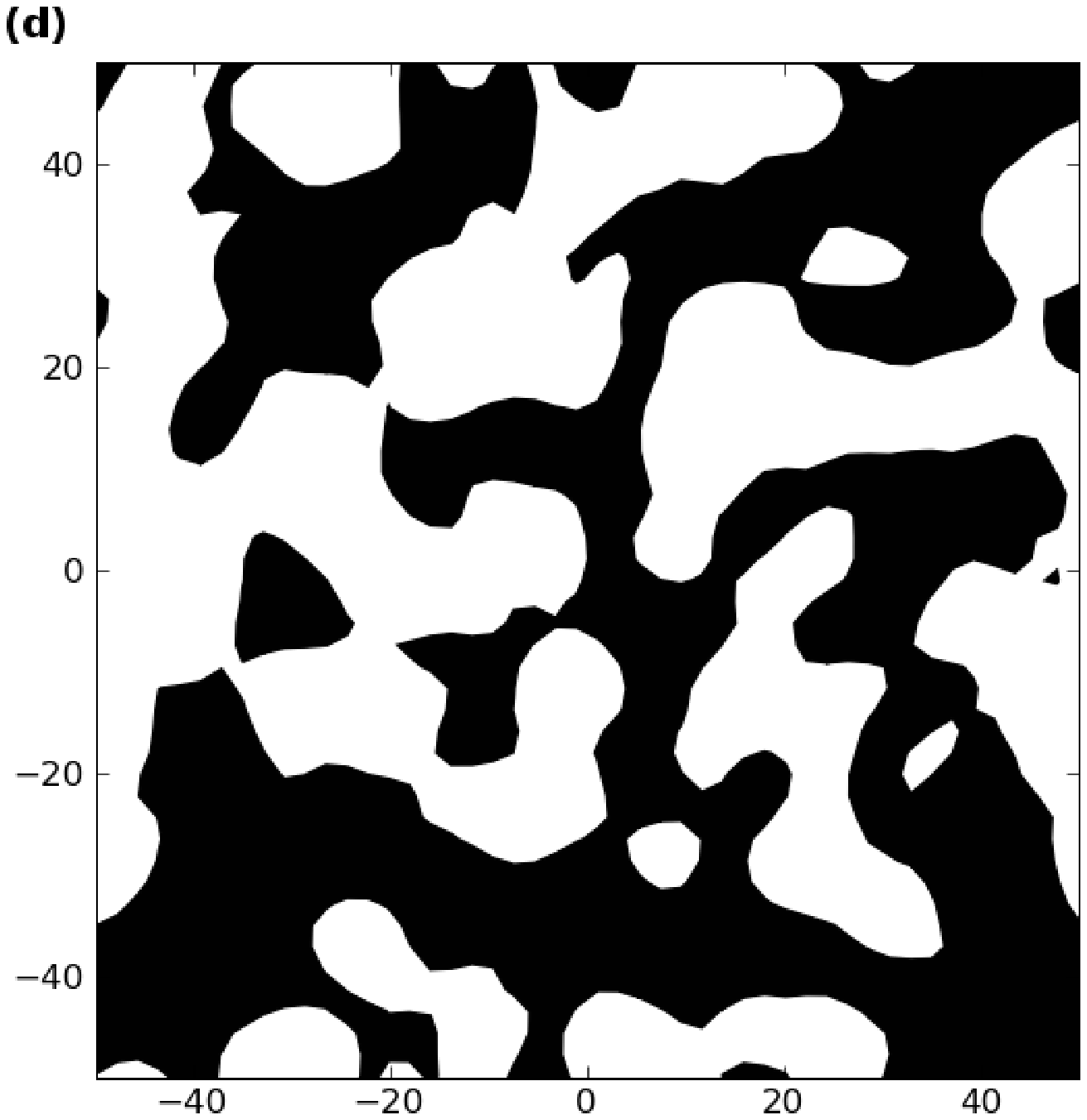}
\includegraphics*[width = 0.4\textwidth]{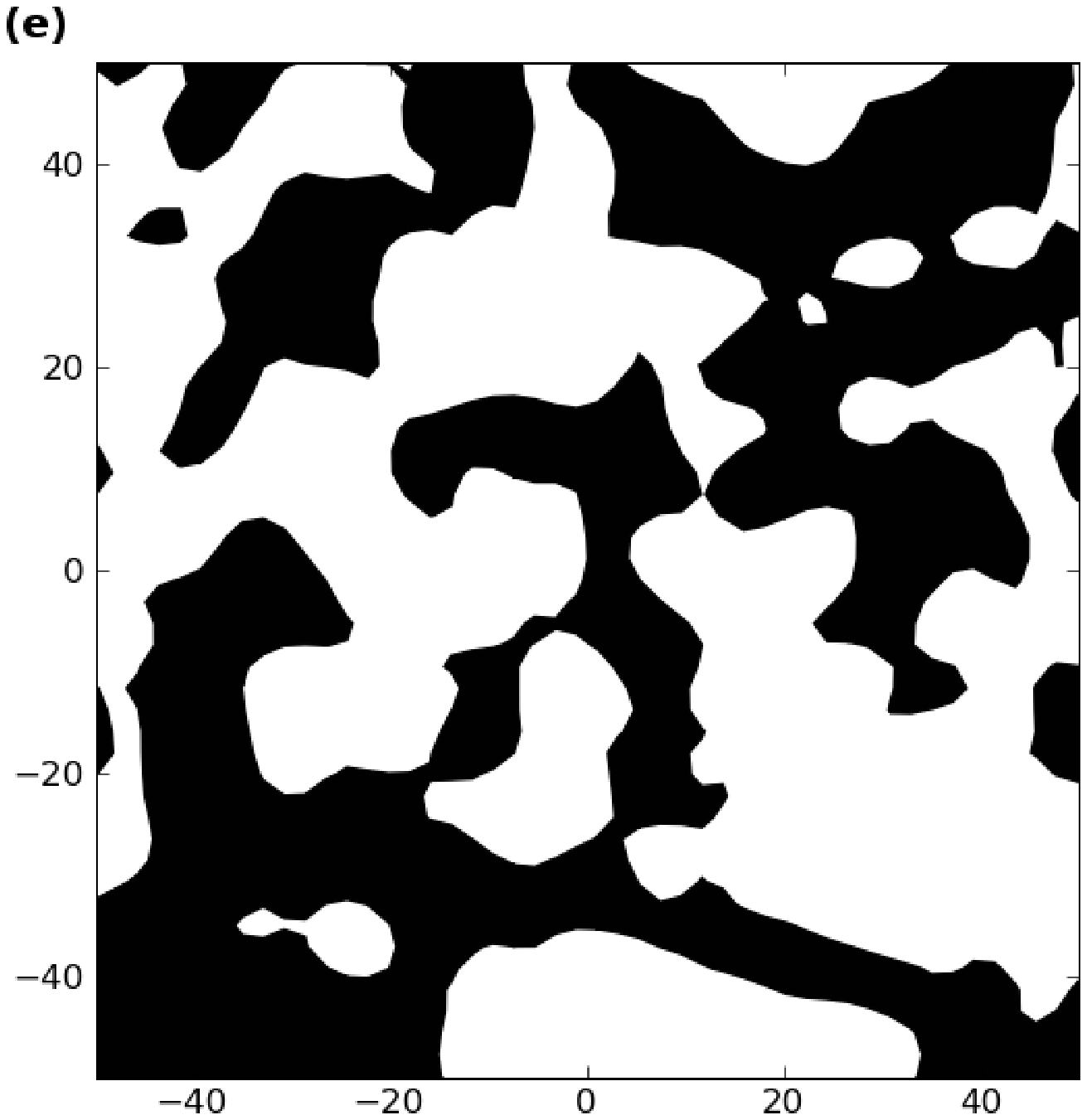}
\includegraphics*[width = 0.4\textwidth]{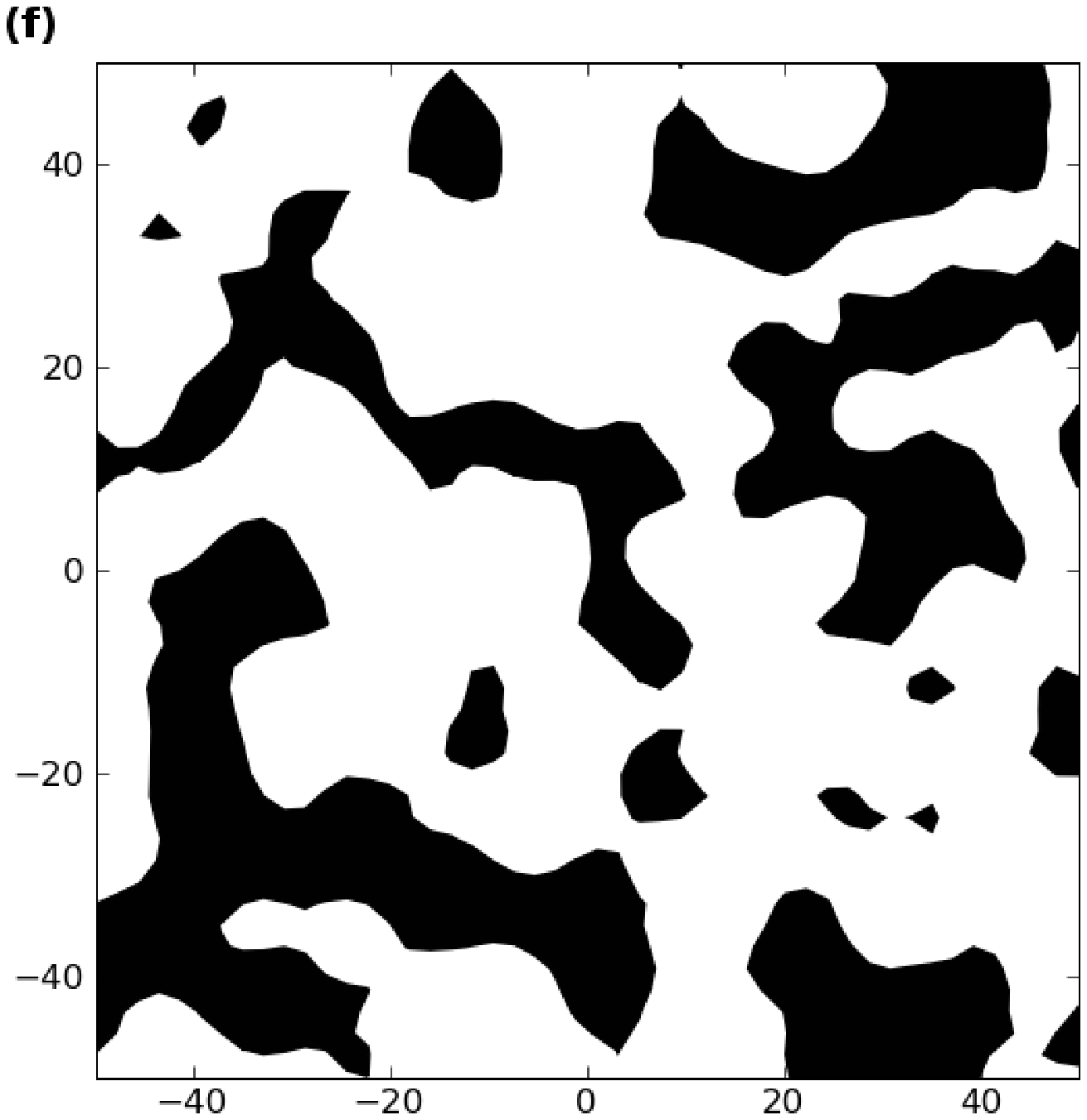}
\caption{Shear modulus divided in rigid ($c_{1}\ge\overline{c_{1}}$, black) and soft zones ($c_{1}\le\overline{c_{1}}$, white) for different macroscopic strains. Figures (a) to (f) correspond to a macroscopic strain of (a)$2.5\%$, (b)$2.55\%$, (c)$2.65\%$,(d)$2.7\%$,(e)$3.25\%$ and (f)$3.4\%$.}
\label{fig13}
\end{figure*}

In order to describe the relation between the local elasticity of the material and its  dynamics, one can try to quantify the connection between the local domains presenting small local modulus and an increased mobility of the particles in these domains. A vast literature has grown in the last five years on the  connection between static structural properties and dynamical heterogeneities in glass formers. While different approaches have partially failed to support such a link between local structure (such as local free volume, local inherent state potential energy, defects, Voronoi tessellation, local stress or strain...) and dynamics, Harrowell et al \cite{Widmer-Cooper2006} have recently shown that the spatially heterogeneous `local Debye-Waller (DW) factor' (defined as the mean-squared vibration amplitude of a molecule over a time of approximately 10 periods of oscillation of this molecule) in a two-dimensional glass forming mixture could be mapped perfectly on the locally measured dynamical propensity that relates to the long term dynamical heterogeneities in the material. Berthier et al \cite{Berthier2007} pursued this discussion showing how the influence of structure on dynamics is much stronger on large length scales than on shorter ones, and that the choice of the coarse-graining scale in  the structure-dynamics problem is crucial. Here we make connection with this literature and claim that the local order parameter $c_{1}$ is a good candidate to establish a relation between  structure and dynamics. One can understand this assertion by the fact that $c_{1}$ and the DW factor contain a similar physical information in probing the local stiffness of the material. Of course in a quasistatic deformation one can not simply measure a local DW factor on short time scales and the order parameter $c_{1}$ is a good measure of the stiffness of a region. Following Cooper and Harrowell in \cite{Widmer-Cooper2006} we define a quasistatic analogue of the dynamic propensity as $\langle\left[{\bf r}_{i}(\epsilon)-{\bf r}_{i}(0)\right]^{2}\rangle$ where ${\bf r}_{i}(\epsilon)-{\bf r}_{i}(0)$ is the displacement without the affine contribution due the macroscopic strain $\epsilon$.
Unlike the original definition of the dynamical propensity in \cite{Widmer-Cooper2006} the average is taken here over all the particles in a given range of the order parameter $c_{1}$ and not over an iso-configurational ensemble of N-particle trajectories. Figure \ref{fig14} shows that the order parameter $c_{1}$ is indeed related to the long term propensity and that soft regions ($c_{1}\le\overline{c_{1}}$) present an increased non-affine mobility in comparison with more rigid regions of the material ($c_{1}\ge\overline{c_{1}} $).

\begin{figure*}
\centering
\includegraphics*[width = 0.4\textwidth]{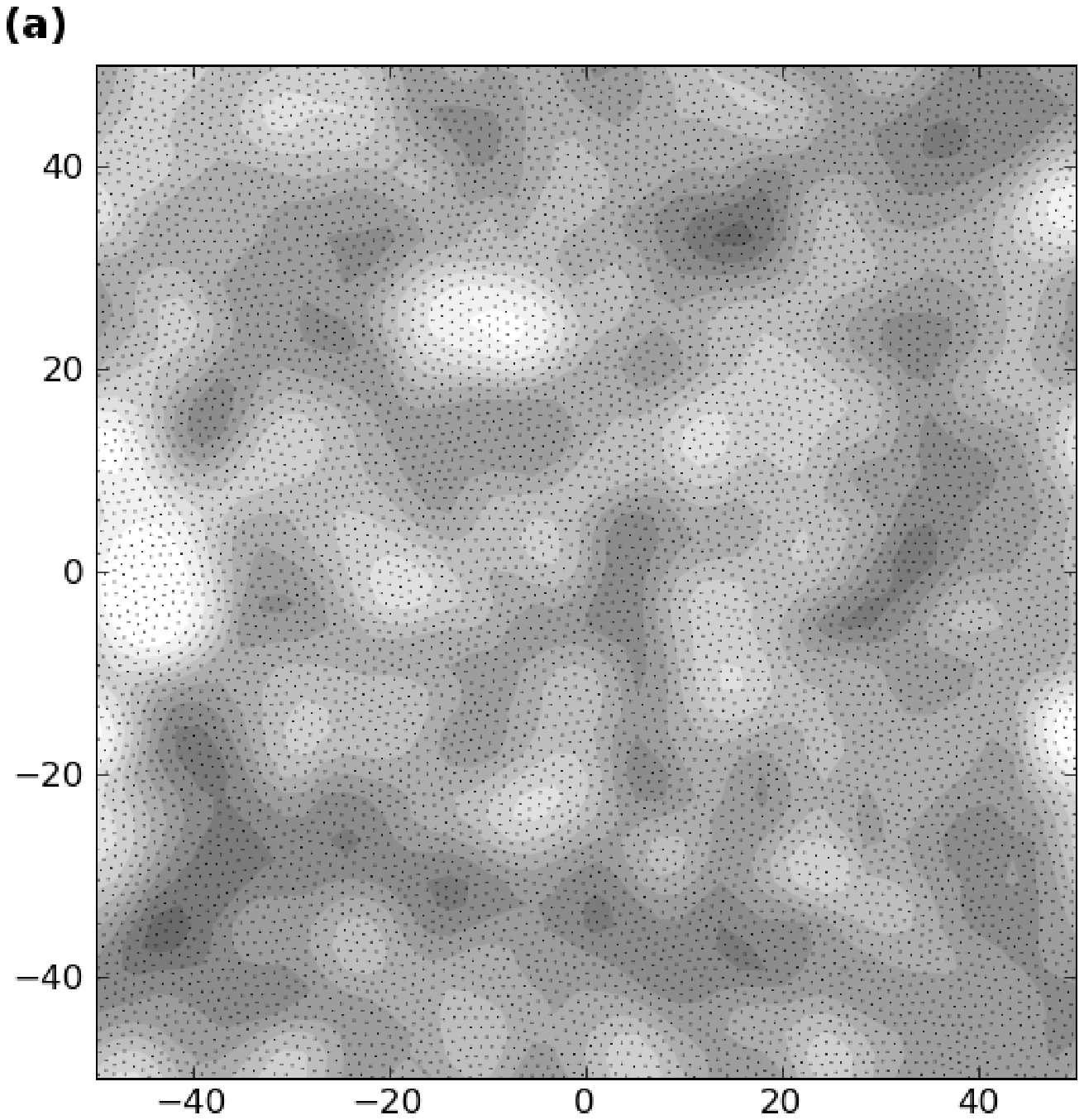}
\includegraphics*[width = 0.4\textwidth]{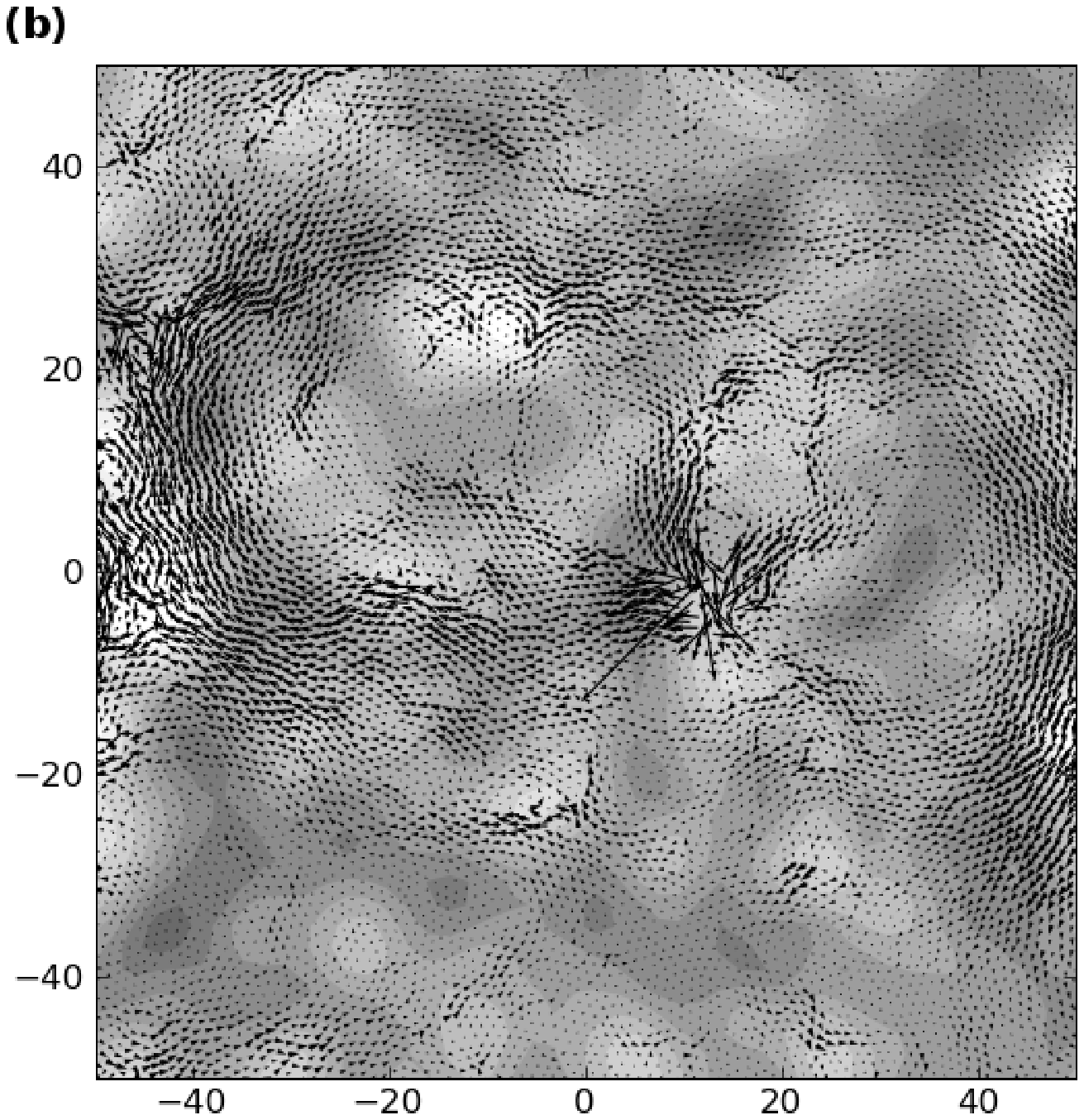}
\includegraphics*[width = 0.4\textwidth]{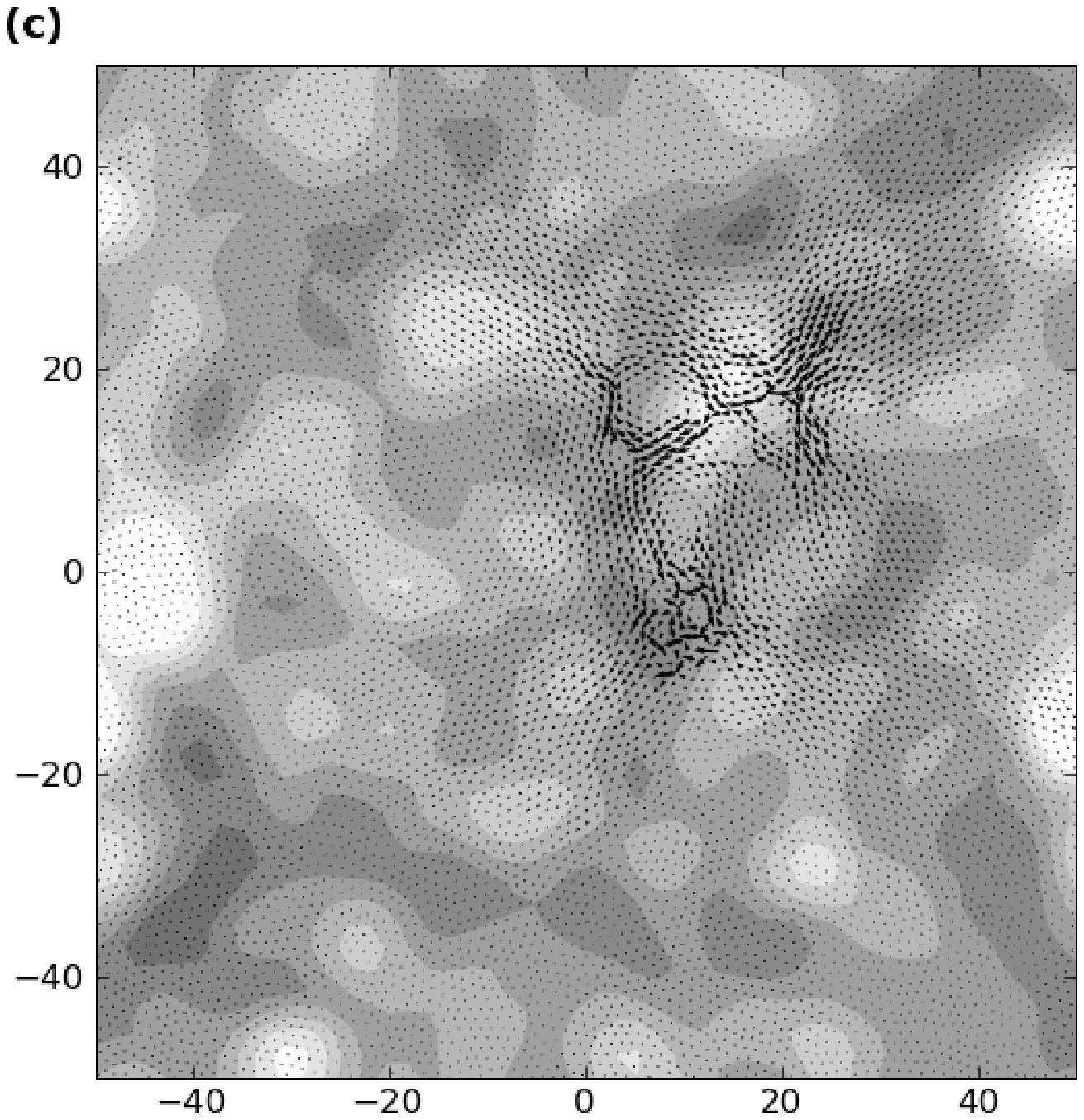}
\includegraphics*[width = 0.4\textwidth]{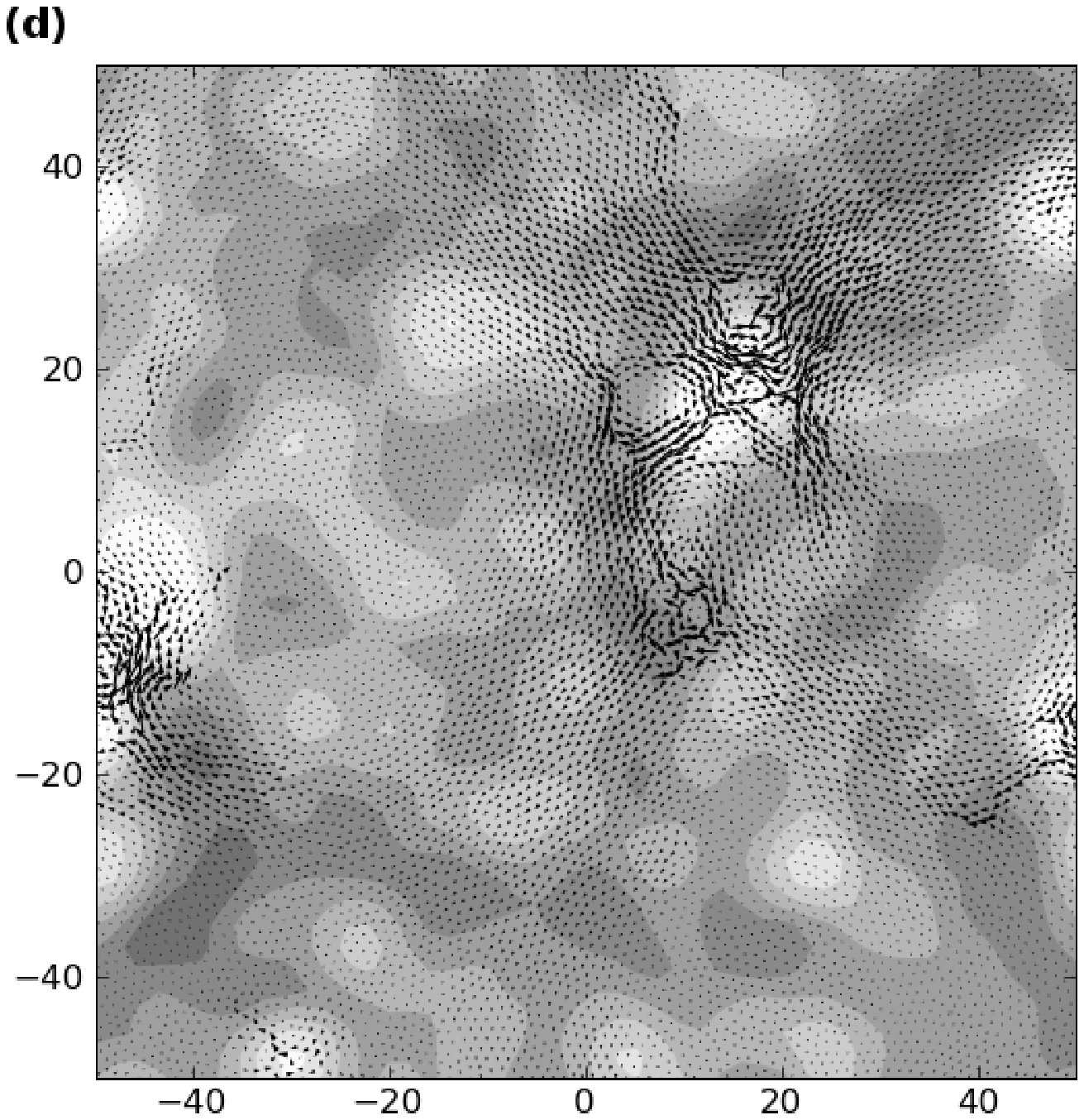}
\includegraphics*[width = 0.4\textwidth]{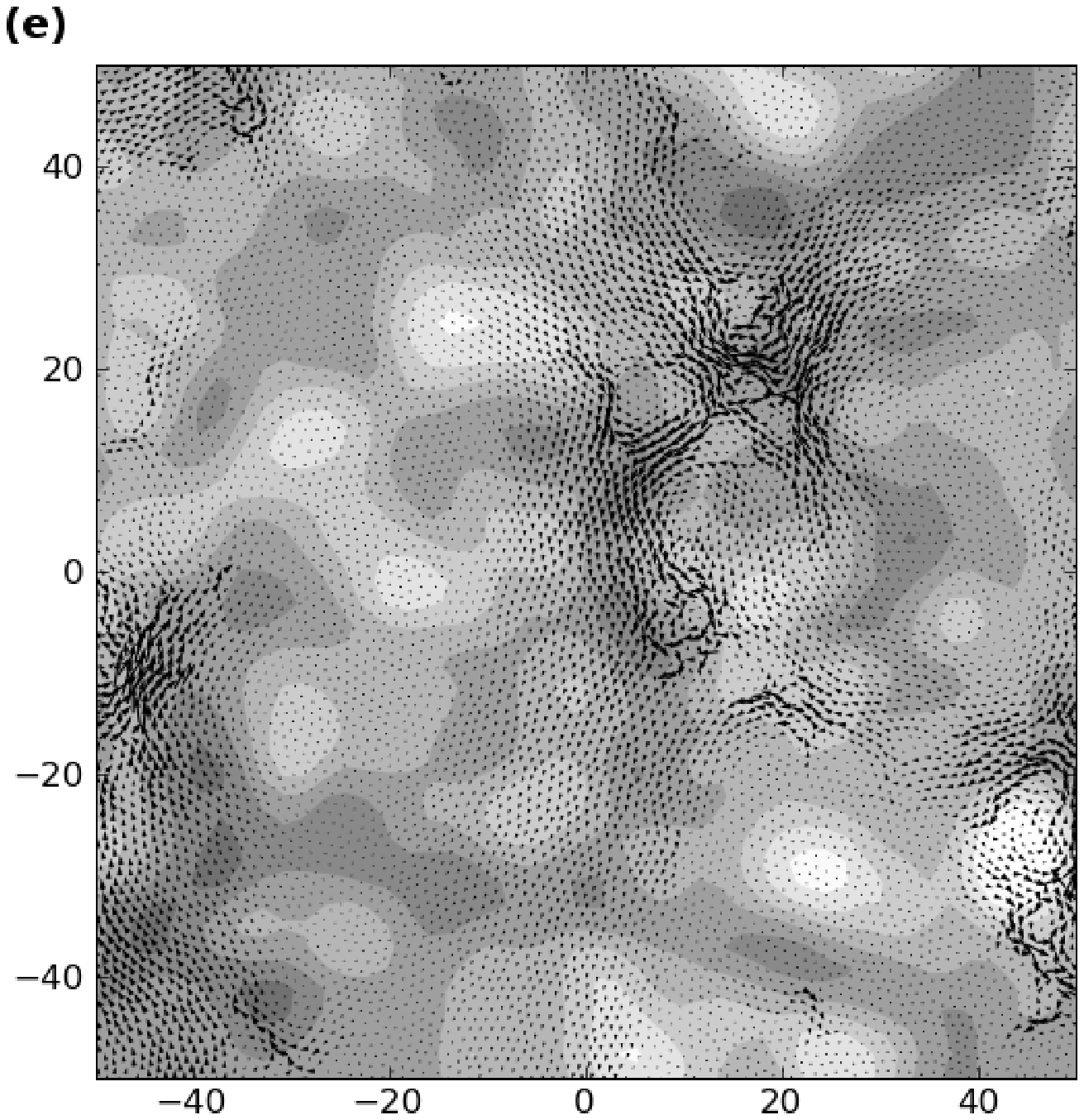}
\includegraphics*[width = 0.4\textwidth]{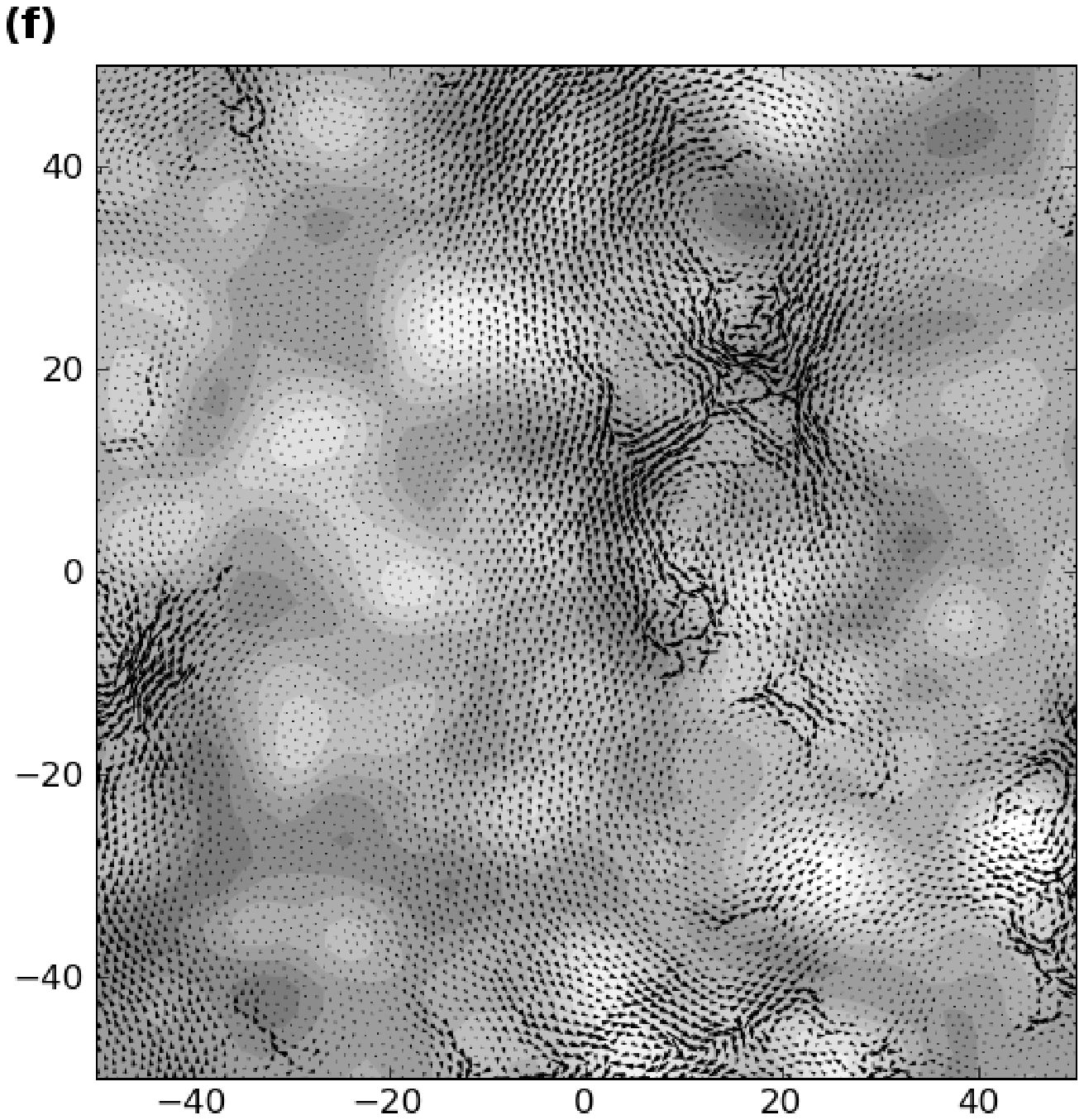}
\caption{We represent the local map of the shear modulus at the same macroscopic strain values as in figure 13. These values correspond to the onset of plastic rearrangement of the material. These maps are superimposed with the non-affine displacement accumulated from figure (a). Figures (a) to (f) correspond to a macroscopic strain of (a)$2.5\%$, (b)$2.55\%$, (c)$2.65\%$,(d)$2.7\%$,(e)$3.25\%$ and (f)$3.4\%$. Note that the non-affine field is multiplied by a factor 300 on fig (b) to illustrate the very strong correlation of the elastic non-affine field with the elasticity map for small incremental strain intervals (here $0.05\%$).}
\label{fig15}
\end{figure*}

\begin{figure}[!hbtp]
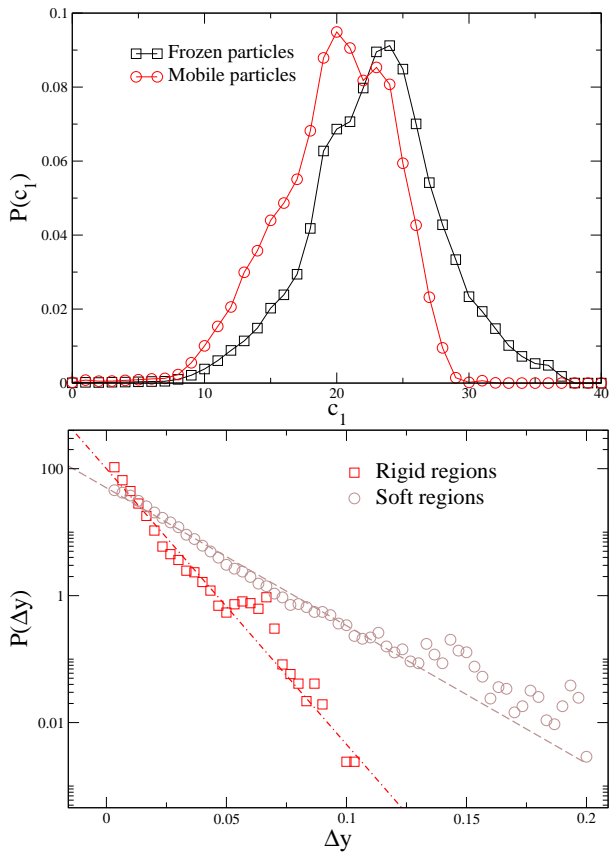

\begin{center}
\includegraphics*[width = 8cm]{figure16a.eps}
\includegraphics*[width = 8cm]{figure16b.eps}
\caption{On the left the distribution of the local shear modulus for mobile and immobile particles. On the right distribution of the nonaffine field during a total strain of $1\%$ for rigid and soft regions.}
\label{fig16}
\end{center}
\end{figure}

The relation between mobility (plasticity) and low shear modulus is illustrated in figure \ref{fig15} by looking at how the spatial distribution of these two quantities are mapping onto each other.
In this figure, the cumulative non-affine displacement field (that is essentially irreversible) appears to nucleate from the initial reference state near the soft zones of the material and to grow in a cooperative manner up to the point where the material fails macroscopically forming a vertical shear band across the sample. In figure \ref{fig16} we distinguish between mobile and immobile particles (Fig. \ref{fig16} left), and between soft and rigid zones (Fig. \ref{fig16} right). The mobile and frozen particles are identified somehow artificially by the amplitude of the transverse nonaffine displacement: $\Delta y\ge 0.02$ for a total strain of $1\%$ and $\Delta y\le0.02$ respectively. The soft zones are identified by $c_{1}\le\overline{c_{1}}$ and the rigid zones by $c_{1}\ge\overline{c_{1}}$. In figure \ref{fig16}, we plot the distribution of shear modulus associated with each group of particles (mobile and frozen) and the distribution of nonaffine displacements for the rigid and soft portion of the sample. Figure \ref{fig16} confirms the visual impression of figure \ref{fig15} that most of the displacement is concentrated in the soft regions of the material and conversely that mobile particles are located in soft zones.

\begin{figure}[!hbtp]
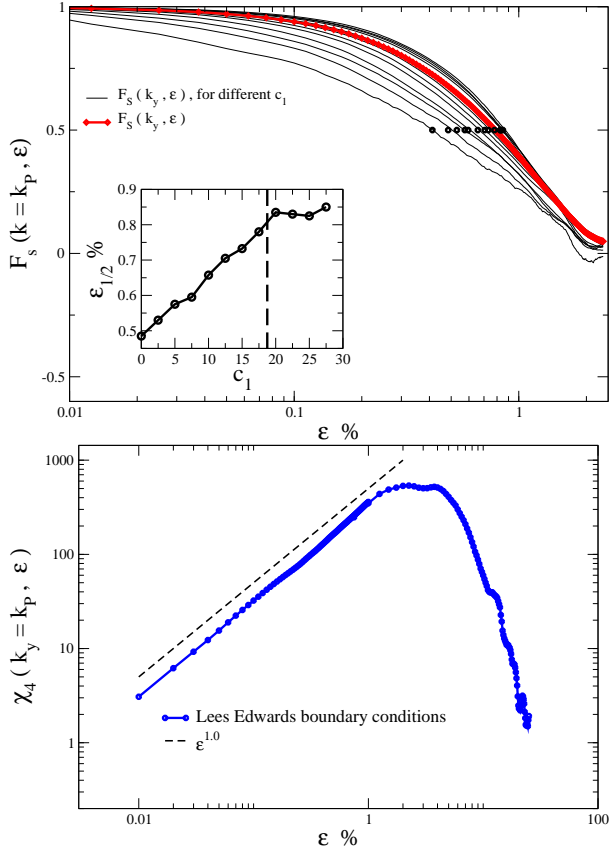

\begin{center}
\includegraphics*[width = 8cm]{figure17a.eps}
\includegraphics*[width = 8cm]{figure17b.eps}
\caption{On the left the self intermediate scattering function (SISF) $F_{s}({\bf k},\epsilon)$ is plotted for \textbf{k}$=k_{x}=k_{P}$ (squares) and \textbf{k}$=k_{y}=k_{P}$ (diamonds) and for regions of different softness $c_{1}$ (thin lines, from bottom to top the rigidity $c_{1}$ is increased). Inset in the left: the relaxation strain is shown as a function of the softness at the origin. On the right panel the 4 point correlation function is represented for rigid (squares) and Lees-Edwards boundary conditions (circles). For small strain increments the linear behavior is highlighted.}
\label{fig17}
\end{center}
\end{figure}

We showed that most of the displacement occurs in soft zones. As seen in figure  \ref{fig15} the dynamics in the sheared glass is not trivial with regions that concentrate most of the non-affine displacement field and others that remain quiescent. The appearance of bursts of mobility seems therefore strongly dictated by the underlying heterogeneous elastic structure of the material and one cannot understand cooperative dynamics in the glass without considering this underlying structure. We would like now to address the question of the degree of cooperativity of this mobility field in the material and its relation with the local elasticity map. In the literature the dynamical heterogeneity of ageing (\cite{Toninelli2005}) or sheared (\cite{Dauchot2005},\cite{Tsamados2008},\cite{Furukawa2009}) glassy systems is commonly quantified by a four-point correlation function defined as:
\begin{equation}
\chi_{4}({\bf k},\epsilon)=\frac{1}{N}\left[\langle F_{s}({\bf k},\epsilon)^{2}\rangle-\langle F_{s}({\bf k},\epsilon)\rangle^{2}\right]
\end{equation}
where $F_{s}({\bf k},\epsilon)$ is the self intermediate scattering function defined by:
\begin{equation}
F_{s}({\bf k},\epsilon)=\sum_{i} cos\left({\bf k}.({\bf r}_{i}(\epsilon)-{\bf r}_{i}(0))\right).
\end{equation}
It is important to note here that the symmetry of the mechanical deformation introduces an anisotropy in k-space in the relaxation of the self intermediate scattering function (SISF) $F_{s}({\bf k},\epsilon)$. Typically one has at the first peak ($k_{P}$) of the static structure factor a relaxation strain of $F_{s}(k_{x}=k_{P},\epsilon)$ of about $0.2\%$ while for $F_{s}(k_{y}=k_{P},\epsilon)$ the relaxation strain is of about $1\%$ (see figure \ref{fig17}(a)). This difference can be attributed to the formation of shear bands in our model glass, preferentially along the x axis, therefore increasing the mobility along this axis. These results seem in contradiction with recently reported similar studies in a model binary supercooled liquid, where an isotropic relaxation is reported for $F_{s}({\bf k},\epsilon)$. It would be interesting to clarify, if these discrepancies could be attributed to the thermal agitation present in \cite{Furukawa2009} and absent in our athermal simulations. Also in figure \ref{fig17}(a) in order to relate mobility and structure we calculate the SISF for different region of the material according to their rigidity. One sees in the inset of figure \ref{fig17}(a) that the relaxation strain associated with each SISF grows linearly with the local shear modulus below the average shear modulus ($c_1\leq \overline{c_1}\approx 18$) and then reaches a plateau at a value of about $\epsilon_{relax}\sim0.85\%$. Again this provides a clear evidence of the connection between the structural order parameter $c_{1}$ and the dynamical response of the material. It cuts the sample into soft ($c_1\leq\overline{c_1}$) and rigid ($c_1 > \overline{c_1}$) zones.

In figure \ref{fig17}(b) we see in $\chi_{4}(k_{y}=k_{P},t)$ that the number of particles evolving in a cooperative way grows linearly with strain from zero to $\sim500$ particles for LE boundary conditions and to $\sim100$ for rigid boundary conditions.
It is indeed system size dependent and evolves as $N.f(\xi/L)$, as shown in \cite{Lemaitre2009,Krasa,Tsamados2009}.

The maximum cooperativity is achieved at a strain of $\epsilon\sim2\%$ in the LE case as well as in the rigid walls case. Theoretical predictions concerning the four-point correlation function are reported in (\cite{Toninelli2005}) where the authors focus their studies on static supercooled liquids near the glass transition temperature. Whether or not one can identify the dynamical correlation length scale to the typical spatial extent of the soft zones of the material remains unclear from this analysis and requires further studies. We note that the typical strain at which the cooperativity $\chi_{4}(k_{y}=k_{P},t)$ is maximal ($\epsilon_{max}\sim2\%$) does not correspond exactly to the structural $\alpha$-relaxation strain $\epsilon_{\alpha} < 1\%$ defined as $F_{s}(k_{y}=k_{P},\epsilon_{\alpha})=1/e$ calculated for the same wave vector, but the order of magnitude is actually the same for the systems studied here.

\subsection{Predicting plastic activity}
\label{plasticity}

To understand the dynamics in the soft phase and more generally the rheology (or mechanical response) of the material one would like to understand what first triggers the nucleation source points (local plastic events) at some specific locations and second how these local rearrangements interact in a cooperative manner.

In the previous section  we have analyzed the coupling between the elasticity map and the nonaffine field in the material and claimed that, for sufficiently large strain,  the non-affine field (for exemple shown in figure \ref{fig15}) is essentially irreversible (plastic). We checked this assertion by comparing for each particle in the system the total nonaffine displacement and the purely irreversible displacement. The irreversible displacement field is defined as the residual displacement field resulting when after each macroscopic  elementary strain increment $\delta\epsilon\sim5.10^{-5}$ (in figure \ref{fig1}) the `virtual' reverse shear $-\delta\epsilon$ is applied on the system. For a purely reversible deformation this field should cancel exactly, but here one observes that, for most incremental deformation steps, a non-vanishing residual irreversible displacement field is present. With this numerical protocol we can extract for each particle i of the system the purely irreversible displacement ${\bf \Delta r^{i}_{irrev}}$ from the total nonaffine displacement ${\bf \Delta r^{i}_{na}}$. Hence for each particle i the total nonaffine displacement over a macroscopic strain $\Delta\epsilon=\sum_{n}\delta\epsilon_{n}$ can be decomposed into ${\bf \Delta r^{i}_{na}}(\Delta\epsilon)=\sum_{n}{\bf \Delta r^{i}_{irrev}}(\delta\epsilon_{n})+\sum_{n}{\bf \Delta r^{i}_{rev}}(\delta\epsilon_{n})$. The relative error $\Vert{\bf \Delta r^{i}_{na}}(\Delta\epsilon)-{\bf \Delta r^{i}_{irrev}}(\Delta\epsilon)\Vert/\Vert{\bf \Delta r^{i}_{na}}(\Delta\epsilon)\Vert$ averaged over all particle is obtained and being less than $5\%$ confirms our assumption that the nonaffine displacement is dominated by an irreversible plastic contribution. Based on this observation we describe here the link between plasticity in the material and the local elasticity map. 
To obtain this information we study here the dynamics of the local rigidity $c_{1}$ calculated on each particle over a strain range of $10\%$.

\begin{figure}[!hbtp]
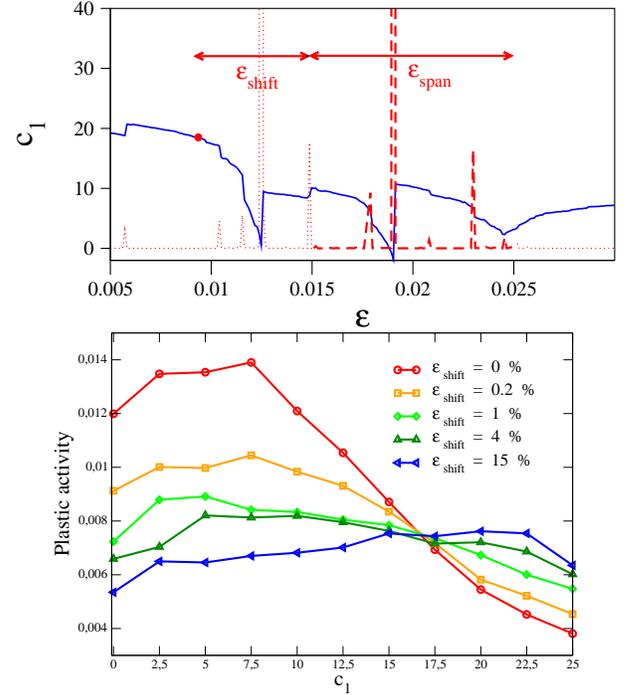

\begin{center}
\includegraphics*[width = 8cm]{figure18a.eps}
\includegraphics*[width = 7cm]{figure18b.eps}
\caption{(a): Evolution of $c_{1}$ of a particle in a plastic zone. The dotted line represents the parameter $D_{BF}$ that marks the local deviation from affinity. $\epsilon_{shift}$ and $\epsilon_{span}$ are defined for the point marked in red. (b): cumulative plastic activity for $\epsilon_{span}\sim0.5\%$ and for different values of $\epsilon_{shift}$.}
\label{fig18}
\end{center}
\end{figure}

\begin{figure}[!hbtp]
\begin{center}
\includegraphics*[width = 0.3\textwidth,angle=-90]{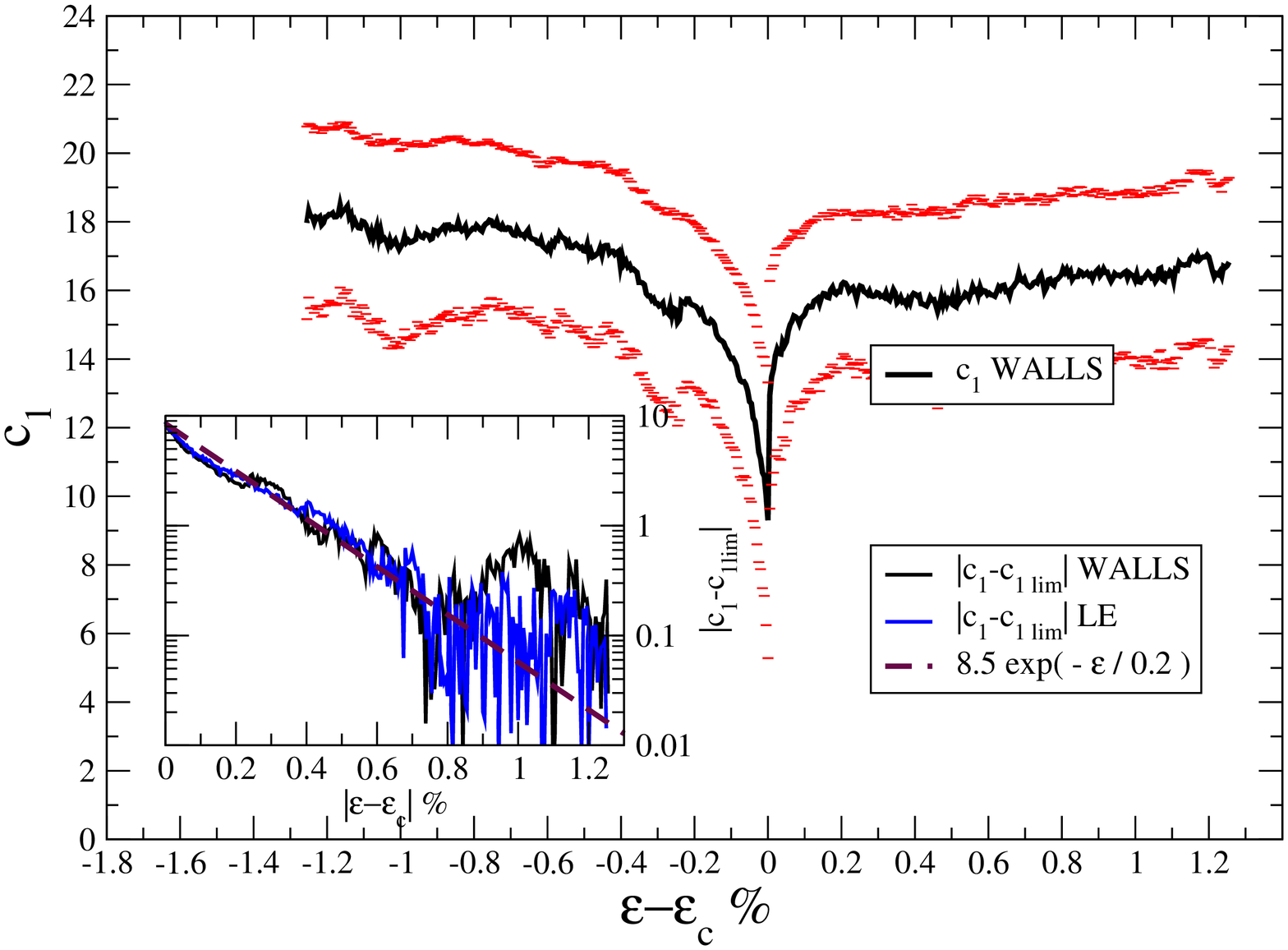}
\includegraphics*[width = 0.3\textwidth,angle=-90]{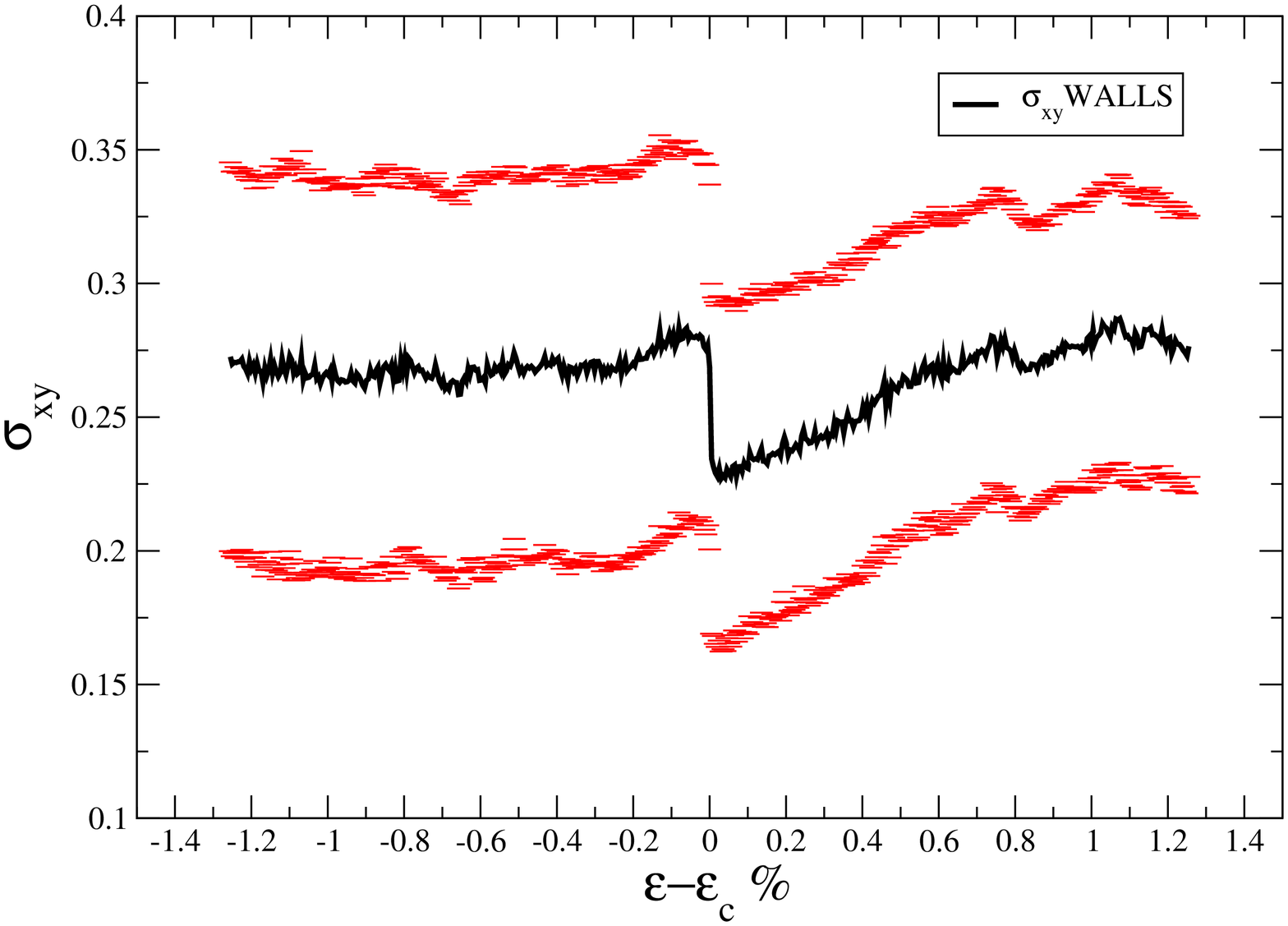}
\includegraphics*[width = 0.3\textwidth,angle=-90]{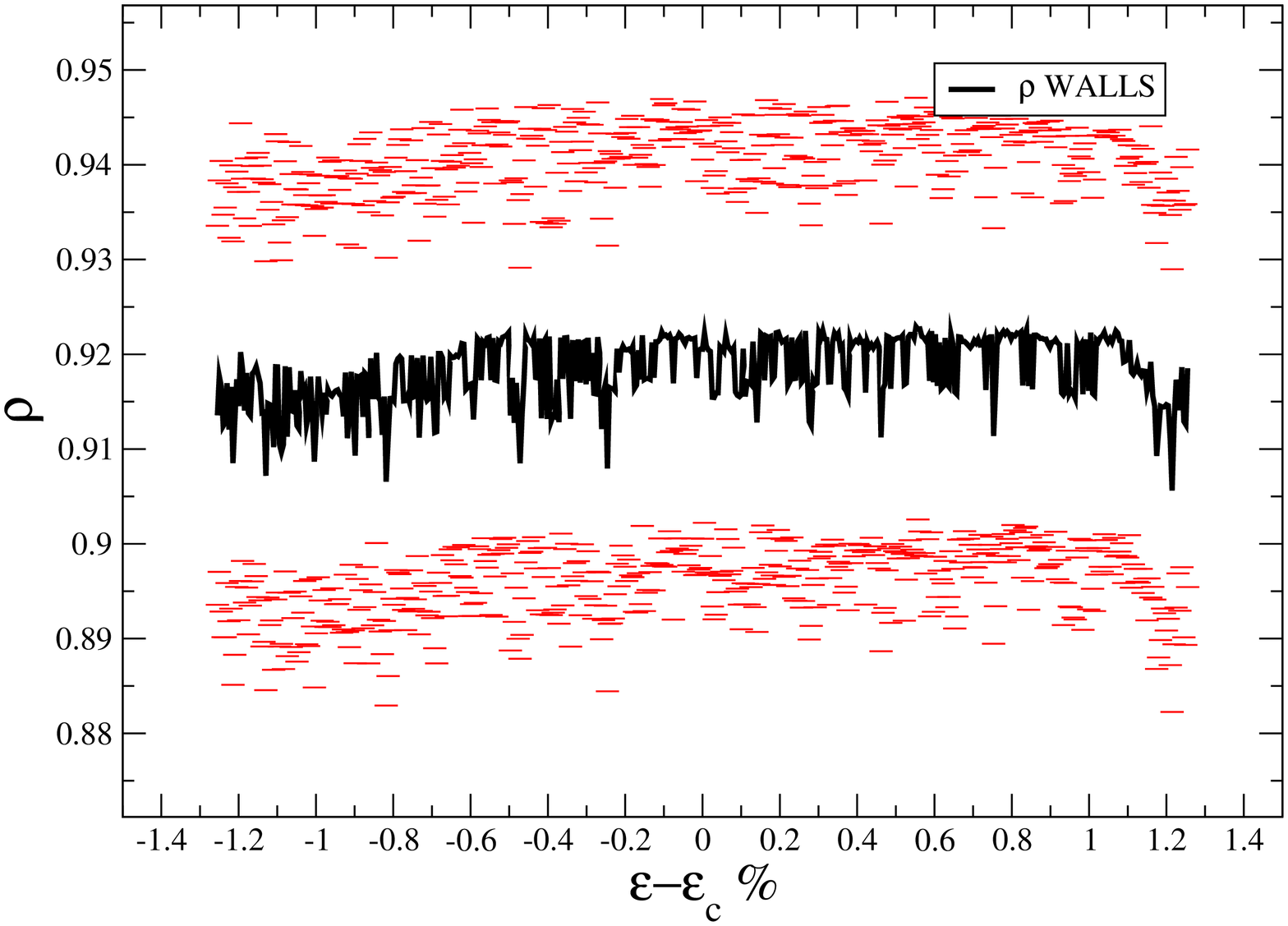}
\includegraphics*[width = 0.3\textwidth,angle=-90]{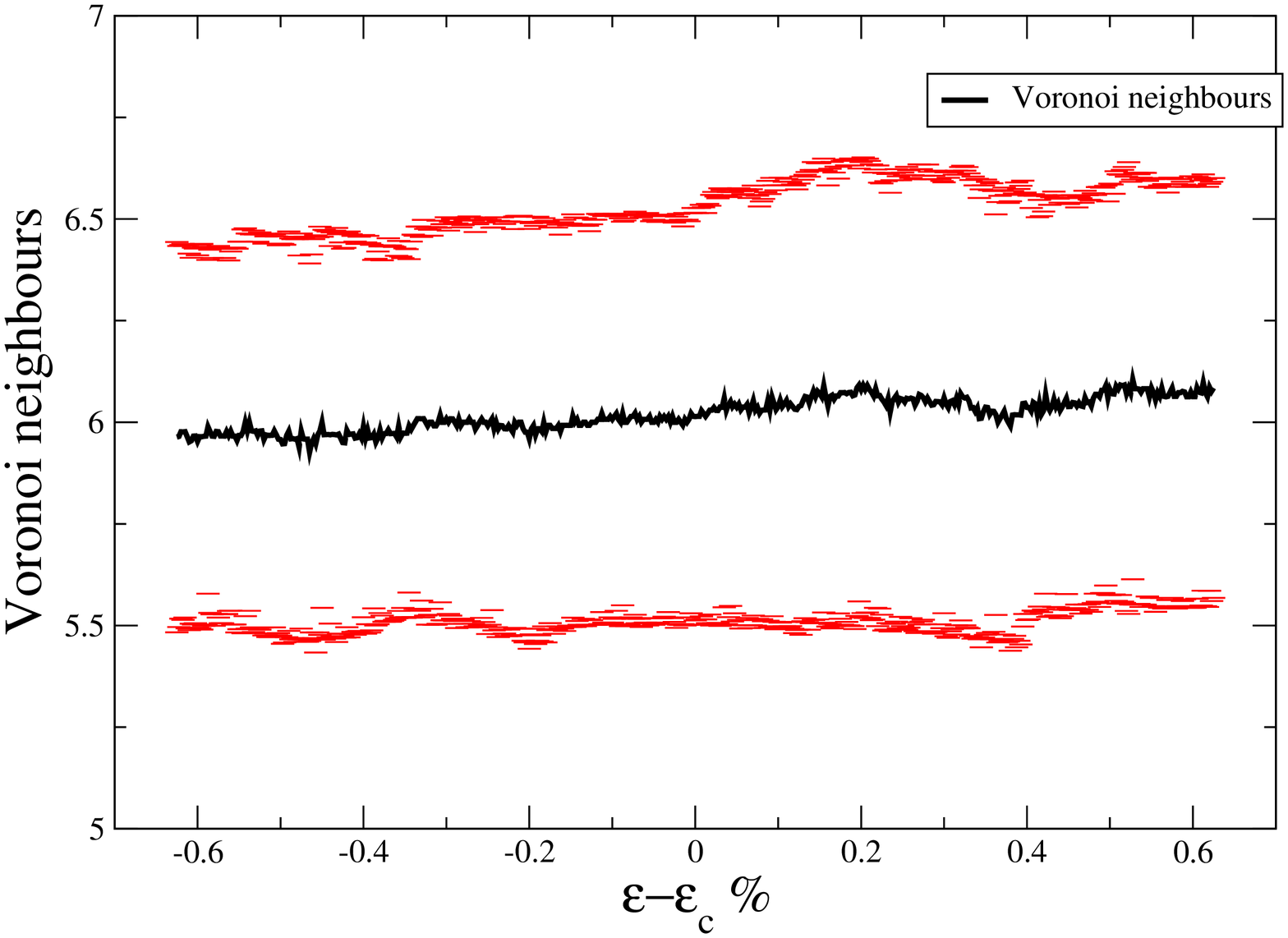}
\caption{Evolution of different fields around a plastic event and averaged over all plastic events. (a) local shear modulus, (b) local shear stress, (c) local density and (d) number of Voronoi neighbors.}
\label{fig19}
\end{center}
\end{figure}

Figure \ref{fig18} illustrates the evolution of the coarse-grained modulus $c_{1}$ on a particle that experiences plastic activity (i.e. that rearranges over the strain range). We have superimposed the quantity $D_{BF}$ defined in \cite{Falk1998}, which evaluates the degree of local deviation from affinity. This parameter was found to very accurately distinguish between plastic like zone ($D_{BF}\sim1$) and normal elastic zones ($D_{BF}<10^{-4}$). Figure \ref{fig19}(a) shows the average behavior of the same quantity $c_1$, averaged for all plastic events. The result which is typical of the dynamics of $c_{1}$ shows that before a plastic event occurs on the site the modulus $c_1$ decreases over a typical strain interval of about $0.2\%$ to become zero or even negative at the plastic irreversible event, where the non-affine displacement becomes important ($D_{BF}\sim1$). Then the local structure is relaxed and the local modulus gets a higher value ($c_{1}\sim 18$ after the event).
Note that this value $c_1\sim 18$ is smaller than the macroscopic value for $2\mu$ but corresponds to the average value $\overline c_1$ of the shear modulus at the scale $W=5$ of description. Just before a plastic event, the local shear modulus is very low $c_1<<\overline{c_1}$.
The average decrease of $c_1$ before a plastic event is fitted approximately by an exponential decay to its limit value.
We have no explanation for the moment for this exponential fit. It shows a characteristic strain $\sim 0.2\%$.

It is interesting to compare in our system this order parameter with other possible predictors of plasticity introduced in the literature. These parameters are the local stress \cite{Bulatov1994,Baret2002,Picard2004}, the local deformation strain \cite{Sollich1997,Lemaitre2007a}, the local free volume \cite{Falk1998,Nemilov2006,Dyre1996}, coordinance defects \cite{Talati2009,Downton2008} as well as other local criteria derived from macroscopic mechanics such as Tresca local yield criterion or a local Mohr-Coulomb \cite{Rottler2001,Schuh2003}. In figure \ref{fig19} we have summarized the evolution of some of these fields measured locally in our model glass at the sites that undergo plastic rearrangement before, during and after the relaxation takes place. The curves are averaged over all plastic events over the strain studied ($\approx 1000$ events).
It shows that the density and the inverse compressibility are not affected by the plastic event. The shear stress is affected since it decreases suddenly after the plastic event occurred; but before the plastic event, the variation of shear stress is very small, and on a very small strain-range.
Thus the conclusion appearing from the above analysis is that the local shear modulus $c_1$ is the best criterion in our Lennard-Jones glasses, to identify zones that are about to rearrange.
\\

In order to identify the effect of a very low initial local shear modulus $c_1<<\overline{c_1}$ on the plastic activity over a larger strain span, we have calculated the shifted plastic activity $A(i,\epsilon_{macro};{\epsilon_{shift},\epsilon_{span}}$) measured at each site $i$ and every step in the macroscopic strain $\epsilon_{macro}$ and defined as the integral over a strain range $\epsilon_{span}$ of typically a few $\%$ of the local quantity $D_{BF}$ starting from a shifted value of the macroscopic strain $\epsilon_{macro}+\epsilon_{shift}$ from the actual macroscopic total strain where $c_{1}$ is measured. Note that $\epsilon_{shift}$ and $\epsilon_{span}$ are adjustable parameters.

In figure \ref{fig18} we illustrate the meaning of this two parameters. $\epsilon_{span}$ is the range over which the plastic activity is recorded, in the limiting case of $\epsilon_{span} \rightarrow \infty$ one should obtain the average activity of the glass former. $\epsilon_{shift}$ allows to substract a systematic bias associated with the conditional probability to have an increased plastic activity for low initial values of $c_{1}$. In figure \ref{fig18} we also plot the correlation between the average activity and $c_{1}$ for different $\epsilon_{span}$. The plastic activity is a number that counts the number of significant plastic events. It is incremented by 1, as soon as $D_{BF} > 0.2$. It appears that the probability to encounter a plastic event is larger for originally soft regions ($c_1<\overline{c_1}$). Moreover, the predictive character of the structural softness of the material on the subsequent plastic activity of the glass former holds even for relatively large $\epsilon_{shift}\sim 4 \%$.

\section{Conclusion and perspectives}

We have shown in this paper that the elastic response of a 2D Lennard-Jones glass at very low temperature is heterogeneous. We have characterized the local elastic response by an extensive study of the local elastic moduli as a function of the coarse-graining scale. We have shown that at a very small scale ($W<5$) the elastic response deviates from Hookes's law, and that on an intermediate scale ($5<W<20$) the system is not homogeneous and not isotropic, but becomes approximately homogeneous beyond this scale. However, it is difficult to identify a well defined characteristic scale, because all the quantities probed in this case are power-law dependent of the coarse-graining scale. By considering the spatial oscillations visible in the autocorrelation function of the local shear modulus for $W<10$, we have decided to compute the local moduli at the scale $W=5$ where Hooke's law applies, and where the heterogeneities are clearly identified.

We have shown that these heterogeneities can be related with the local dynamics of the particles. Regions with very low local shear modulus give rise to enhanced non-affine displacements and plastic activity. Moreover, the relaxation of low moduli during plastic deformation of the sample is faster that the relaxation of rigid zones, and the system seems to be decomposed into fast  evolving regions with low local shear modulus ($c_1<\overline{c_1}$), and rigid, frozen regions with  high local shear moduli ($c_1>\overline{c_1}$). The occurrence of large plastic events seems to be associated with a large proportion of soft regions in the system (figure\ref{fig12}).

We have also shown that  the local shear moduli can be used as  a predictive local criterion for plasticity. This is true
on very short strain scales, as 
we have seen in that the lowest local shear modulus evolves in a systematic way before a plastic event, 
with a decrease taking place over a  typical strain of $\epsilon\sim 0.2\%$ before a plastic event. Interestingly, this is also true on larger strain scales,
as a site with an initially low modulus tends to remain plastically active  over long periods of strain, of the order of a few percent. However it is more difficult to identify the type of plastic events in which a site will be involved, which may vary considerably in size from isolated quadrupolar events to elementary shear bands spanning the whole system.  The present work is a first step in order to predict plastic activity, and to relate it with the heterogeneous elastic structure in the system.  Further studies will be needed, in order 
 to make contact with existing models of plasticity and rheology of amorphous systems. In particular, such models \cite{Sollich1997}\cite{Picard2004} do not in general introduce the possibility of heterogeneous elasticity. It will
 be interesting to test their sensitivity to this ingredient, and to see if they are able to reproduce the observed correlation between
 low elastic constants and high plastic activity.


The elastic structure in the system (the set of local elastic moduli) is computed by the way of macroscopic deformations and can result from collective effect. We do not known at the moment, if this measurement can be related with a more local and simpler structural property of the samples studied. In the case of model silicon systems, where local tetrahedral order due to the covalent bound is important, it has been shown~\cite{Talati2009} that plastic activity is related with the occurrence of local coordination defects and unusual atomic environnement. It is not the case in Lennard-Jones glasses. It would be very interesting to see if a criterion based on the lowest local elastic shear modulus would be also valid in other systems, independent on the local directionality of bounds.

The relation between plastic activity and elastic structure opens new possibilities in the theoretical and experimental study of the deformation  of glasses. From a theoretical point of view, the detailed study of the dynamical evolution of local elastic moduli should allow to construct a model as we have already done for the local stress components~\cite{Tsamados2008}, including a criterion for local plastic rearrangement. From an experimental point of view, this study shows that the resolution for the measurement of a local elastic modulus should be less that 10 interatomic distances, in order to include a description of the relevant scales of elastic heterogeneities. New experimental methods have been proposed recently in order to evaluate the deformation at the nanometer scale~\cite{nano}. This study should encourage to continue in this way.


\end{document}